\documentclass[subeqn]{article}
  \usepackage[T1]{fontenc}
  \usepackage{amsmath,amssymb,mathrsfs}
  \usepackage[left=4.1cm, right=4.1cm, top=3cm, bottom=3cm]{geometry}
  \usepackage{enumitem}
\usepackage[toc,page]{appendix}

\usepackage[ansinew]{inputenc}
\usepackage{cases}
\usepackage{cases}
\usepackage{array}
\usepackage{mathenv}
\usepackage{graphicx,color}
\usepackage{eurosym}
\usepackage{amsmath,amsfonts,amssymb,amsthm}
\usepackage{mathrsfs}
\usepackage{wasysym}
\usepackage{oldstyle}
\usepackage[T1]{fontenc}
\usepackage{pb-diagram}
\usepackage{xspace}
\usepackage{hyperref}

  \newtheorem{theorem}{Theorem}
  \newtheorem{proposition}{Proposition}
   \newtheorem{lemma}{Lemma}
  \theoremstyle{remark}
  \newtheorem{remark}{Remark}
   \newtheorem{notation}{Notation}

  \newenvironment{dem}{
\trivlist \item[\hskip \labelsep{\bf\ Proof:}]}{\hfill \makebox[2em]{\hfill{\footnotesize $\Box$}}
\vspace{2ex}}
\title{\textbf{Existence of Ground State of an Electron in the BDF Approximation.}}
\author{Sok J\'er\'emy\\
 Ceremade, UMR 7534, Universit\'e Paris-Dauphine,\\
  Place du Mar\'echal de Lattre de Tassigny,\\
  75775 Paris Cedex 16, France.\\ \\
}%

\newcommand{\ket}[1]{\ensuremath{|#1\rangle}\xspace}
\newcommand{\bra}[1]{\ensuremath{\langle #1|}\xspace}
\newcommand{\psh}[2]{\ensuremath{\langle #1\,,\,#2\rangle}\xspace}

\newcommand{\ssum}{\ensuremath{\displaystyle\sum}}
\newcommand{\dint}{\ensuremath{\displaystyle\int}}
\newcommand{\diint}{\ensuremath{\displaystyle\iint}}
\newcommand{\diiint}{\ensuremath{\displaystyle\iiint}}
\newcommand{\diiiint}{\ensuremath{\displaystyle\iiiint}}

\newcommand{\D}{\ensuremath{\mathcal{D}^0}}
\newcommand{\ee}[1]{\ensuremath{E\left(#1\right)}}

\newcommand{\ed}[1]{\ensuremath{\widetilde{E}\left(#1\right)}}
\newcommand{\PP}{\ensuremath{\mathcal{P}^0_-}}
\newcommand{\PPP}{\ensuremath{\mathcal{P}^0_+}}

\newcommand{\EE}{\ensuremath{\mathcal{E}_{\text{BDF}}^0}}

\newcommand{\g}{\ensuremath{\gamma}}
\newcommand{\G}{\ensuremath{\Gamma}}

\newcommand{\llo}{\ensuremath{\log(\Lambda)}}
\newcommand{\WW}{\ensuremath{b_{\Lambda}}}
\newcommand{\la}{\ensuremath{\lambda}}
\newcommand{\La}{\ensuremath{\Lambda}}

\newcommand{\hl}{\ensuremath{\mathfrak{H}_\Lambda}}

\newcommand{\rr}{\ensuremath{\mathfrak{R}}}
\newcommand{\ph}{\ensuremath{\varphi}}
\newcommand{\unp}{\ensuremath{\underline{\psi}}}
\newcommand{\lpsi}{\ensuremath{\psi_\lambda}}

\newcommand{\dd}{\ensuremath{\mathrm{d}}}
\newcommand{\eps}{\ensuremath{\varepsilon}}
\newcommand{\oo}[1]{\ensuremath{\omega_#1}}
\newcommand{\om}{\ensuremath{\omega}}

\newcommand{\tr}{\ensuremath{\mathrm{Tr}_{0}}}
\newcommand{\ttr}{\ensuremath{\mathrm{Tr}}}

\newcommand{\tigma}{\ensuremath{T_{\sigma}}}
\newcommand{\talpha}{\ensuremath{T_{\boldsymbol{\alpha}}}}

\newcommand{\CC}{\ensuremath{\mathbf{C}^4}}
\newcommand{\RR}{\ensuremath{\mathbf{R}^3}}

\newcommand{\ov}[1]{\ensuremath{\overline{#1}}}
\newcommand{\un}[1]{\ensuremath{\underline{#1}}}
\newcommand{\ww}[1]{\ensuremath{\widehat{#1}}}

\newcommand{\wh}[1]{\ensuremath{\widehat{#1}}}

\newcommand{\nq}[1]{\ensuremath{\lVert#1\rVert_{\mathcal{Q}}}}
\newcommand{\nqu}[1]{\ensuremath{\lVert#1\rVert_{q_1}}}
\newcommand{\nqo}[1]{\ensuremath{\lVert#1\rVert_{q_0}}}
\newcommand{\nqf}[1]{\ensuremath{\lVert#1\rVert_{\mathcal{Q}_f}}}
\newcommand{\nc}[1]{\ensuremath{\lVert#1\rVert_{\mathfrak{C}}}}
\newcommand{\ncu}[1]{\ensuremath{\lVert#1\rVert_{\mathfrak{C}_1}}}
\newcommand{\ncf}[1]{\ensuremath{\lVert#1\rVert_{\mathfrak{C}_f}}}
\newcommand{\nqr}[1]{\ensuremath{\lVert#1\rVert_{\mathcal{R}}}}
\newcommand{\ns}[2]{\ensuremath{\lVert#2\rVert_{\mathfrak{S}_{#1}}}}
\newcommand{\nlp}[2]{\ensuremath{\lVert#2\rVert_{L^{#1}}}}
\newcommand{\nso}[2]{\ensuremath{\lVert#2\rVert_{H^{#1}}}}
\newcommand{\nhi}[1]{\ensuremath{\lVert#1\rVert_{E}}}
\newcommand{\nqq}[1]{\ensuremath{\lVert#1\rVert_{\text{Ex}}}}
\newcommand{\nqkin}[1]{\ensuremath{\lVert#1\rVert_{\text{Kin}}}}
\newcommand{\nb}[1]{\ensuremath{\lVert#1\rVert_{\mathcal{B}}}}
\newcommand{\ncc}[1]{\ensuremath{\lVert#1\rVert_{\mathcal{C}}}}
\newcommand{\nx}[1]{\ensuremath{\lVert#1\rVert_{\mathcal{X}}}}
\newcommand{\nxu}[1]{\ensuremath{\lVert#1\rVert_{\mathcal{X}_1}}}
\newcommand{\nxf}[1]{\ensuremath{\lVert#1\rVert_{\mathcal{X}_f}}}
\begin{document}
%\pageblanche0
\maketitle
%%%%%%%%%%%%%%%%%%%%%%%%%%%%%%
%\romanpagenumbers
%\input premieres.tex
%%%%%%%%%%%%%%%%%%%%%%%%%%%%%%
%\input MESmacros.tex
%%%%%%%%%%%%%%%%%%%%%%%%%%%%%%

\abstract{The Bogoliubov-Dirac-Fock (BDF) model allows to describe relativistic electrons interacting with the Dirac sea. It can be seen as a mean-field approximation of Quantum Electro-dynamics (QED) where photons are neglected. 

This paper treats the case of an electron together with the Dirac sea in the absence of any external field. Such a system is described by its one-body density matrix, an infinite rank, self-adjoint operator which is a compact pertubation of the negative spectral projector of the free Dirac operator. 

The parameters of the model are the coupling constant $\alpha>0$ and the ultraviolet cut-off $\Lambda>0$: we consider the subspace of squared integrable functions made of the functions whose Fourier transform vanishes outside the ball $B(0,\La)$. We prove the existence of minimizers of the BDF-energy under the charge constraint of one electron and no external field provided that $\alpha,\La^{-1}$ and $\alpha\llo$ are sufficiently small. The interpretation is the following: in this regime the electron creates a polarization in the Dirac vacuum which allows it to bind.

We then study the non-relativistic limit of such a system in which the speed of light tends to infinity (or equivalently $\alpha$ tends to zero) with $\alpha\llo$ fixed: after rescaling the electronic solution tends to the Choquard-Pekar ground state.}

\tableofcontents

\section{Introduction}

The relativistic quantum theory of electrons is based on the free Dirac operator 

\noindent $D^0=-i\hbar c \boldsymbol{\alpha}\cdot\nabla +mc^2\beta$. Here $\beta$ and $\alpha_k$ are the $\mathbf{C}^4\times \mathbf{C}^4$ matrices:
\[
\begin{array}{l}
\beta:=\begin{pmatrix} \text{Id}_2 & 0 \\ 0 & -\text{Id}_2 \end{pmatrix},\ \alpha_k=\begin{pmatrix} 0 & \sigma_j \\ \sigma_k & 0\end{pmatrix},
\end{array}
\]
\[
\begin{array}{lll}
\sigma_1=\begin{pmatrix} 0 & 1\\ 1 & 0\end{pmatrix},& \sigma_2=\begin{pmatrix} 0 & -i \\  i & 0\end{pmatrix},& \sigma_3=\begin{pmatrix}1 & 0\\ 0 &-1 \end{pmatrix}.
\end{array}
\]
The free Dirac operator $D^0$ acts on 4-spinors, that is on $\mathfrak{H}=L^2(\mathbf{R}^3,\mathbf{C}^4)$ which is the Hilbert space of one relativistic electron. It is self-adjoint with domain $H^1(\mathbf{R}^3,\mathbf{C}^4)$ and form domain $H^{1/2}(\mathbf{R}^3,\mathbf{C}^4)$. Moreover $(D^0)^2=m^2c^4-\hbar^2c^2\Delta$.
%[
We write:
\begin{equation}
P^0_-=1-P^0_+:=\chi_{(-\infty,0)}(D^0).
\end{equation}
It is well known that its spectrum is $\sigma(D^0)=(-\infty,-mc^2]\cup[mc^2,+\infty)$ leading to difficulties in relativistic quantum mechanics. This operator was introduced by Dirac to describe the energy of a free particle with spin $\tfrac{1}{2}$ (\emph{e.g.} an electron). To explain why electrons with negative energies are not observed, Dirac postulated that all the negative energy states are already occupied by virtual electrons, the so-called Dirac sea. By the Pauli principle, a real electron cannot have a negative energy.%]

We study an approximation of no-photon Quantum Electrodynamics (QED) allowing to describe the behavior of relativistic electrons in an external field interacting with the virtual electrons of the Dirac sea via the electrostatic potential in a mean-field type theory. This so-called Bogoliubov-Dirac-Fock (BDF) model was introduced by Chaix and Iracane \cite{CI} and then studied by Bach \emph{et al.} in \cite{stab}, by Hainzl \emph{et al.} in \cite{HaiSied,ptf,Sc,mf,at} and by Lewin \emph{et al.} in \cite{gs}. In particular in those last papers, the authors are interested in the existence of ground states for this variational model.

Let us sketch how the BDF model is derived from full QED. We use relativistic units $\hbar=c=1$ and set the bare particle mass equal to $1$ and $\alpha=e^2/(4\pi)$.
When photons are neglected, the (formal) Hamiltonian $\mathbb{H}^{\phi}$ of QED acts on the Fock space $\mathcal{F}$ of $\mathfrak{H}$ \cite{Th}:
\begin{equation}
\mathbb{H}^{\phi}=\dint \Psi^*(x)D^0\Psi(x)dx-\dint\phi(x)\rho(x)dx+\frac{\alpha}{2}\diint\frac{\rho(x)\rho(y)}{|x-y|}dxdy.
\end{equation}
Here $\Psi(x)$ is the second-quantized field operator, $\phi$ is the external field and $\rho(x)$ is the density operator:
\begin{equation}
\rho(x)=\displaystyle\sum_{\sigma=1}^4 \big\{\Psi^*(x)_{\sigma}\Psi(x)_{\sigma}-\Psi(x)_{\sigma} \Psi^*(x)_{\sigma}\big\}.
\end{equation}
In the presence of an external density $\nu$, the corresponding external field is 

\noindent $\phi=\alpha \nu*\tfrac{1}{|\cdot|}$. This Hamiltonian is not bounded from below and it is not possible to solve the corresponding minimization problems.

The BDF variational model is obtained from this Hamiltonian by making several approximations. 

The first one consists in restricting the energy to special states in $\mathcal{F}$, the so-called Bogoliubov-Dirac-Fock (BDF) states. They are states $\Omega_P$ which are fully described by their one-body density matrix $P$:
\begin{equation}
P(x,y)_{\sigma,\tau}=\langle \Omega_P | \Psi^*(x)_{\sigma}\Psi(y)_{\tau}|\Omega_P\rangle_{\mathcal{F}}.
\end{equation}
For instance the vacuum state $\Omega_0$ (no electron and no positron) in $\mathcal{F}$ is a BDF state with one-body density matrix $P^0_-$.

One must consider them as an infinite Slater determinant $f_1\wedge f_2\wedge \cdots$ where $(f_i)_{i\ge 1}$ is an orthonormal basis of the range $\text{Ran}(P)$ of $P$.
We will write $P$ instead of $\Omega_P$ for a BDF state: the QED energy can be written in terms of $P$.

%(
In \cite{mf}, Hainzl \emph{et al.} study the corresponding minimization problem of $\mathbb{H}^{0}$ in the space $\mathfrak{H}_{\Lambda}^{L}$  of functions in $L^2([-L/2,L/2)^3,\mathbf{C}^4)$ (with periodic boundary conditions) whose Fourier transform vanishes outside the ball $B(0,\Lambda)$; the constant $\Lambda>0$ is the so-called ultraviolet cut-off. This space has finite dimension and the corresponding Hamiltonian $\mathbb{H}^{0}_L$ is well-defined.

It is then shown that, for each $L>0$ and $0<\alpha<4/\pi$, there exists a minimizer $P_L=\gamma_L+\tfrac{1}{2}$ among BDF states (with energy $E_L(0)$) and that in the thermodynamic limit $L\to+\infty$, $\Gamma_L$ tends in some sense to a self-adjoint, translation-invariant operator $\Gamma_0$ of $\mathfrak{H}_\La$:
\begin{equation}
\mathfrak{H}_\La:=\{f\in \mathfrak{H},\ \text{supp}\,\wh{f}\subset B(0,\La)\}.
\end{equation}
%]
 Moreover $\Gamma_0$ satisfies the following self-consistent equations:
 \begin{equation}\label{eulerlagrange}
 \left\{
 \begin{array}{rl}
  \Gamma_0&=-\dfrac{\text{sign}(\D)}{2},\\
  \D&= D^0-\alpha \dfrac{\Gamma_0(x,y)}{|x-y|}.
 \end{array}
 \right.
 \end{equation}
The operator $\PP=\Gamma_0+\tfrac{1}{2}$ is the orthogonal projection $\chi_{(-\infty,0)}(\D)$ and we write $\PPP=1-\PP$. The operator $\D$ has been previously introduced in \cite{ls} but in another context.

We will now take $\PP$ as reference state. For a one-body density matrix $P$, the formal difference between the QED energies $\mathcal{E}^\nu_{\text{QED}}(P-\tfrac{1}{2})$ and $\mathcal{E}_{\text{QED}}^0(\PP-\tfrac{1}{2})$ gives the following function of $Q:=P-\PP$:

\[
\left\{
\begin{array}{l}
\mathcal{E}_{\text{BDF}}^\nu(Q)=\text{Tr}\Big\{\D (\PP Q \PP +\PPP Q \PPP)\Big\}-\alpha\dint \phi(x)\rho_Q(x)dx+\frac{\alpha}{2}\Big[D(\rho_Q,\rho_Q)-\text{Ex}[Q]\Big],\\
D(\rho_Q,\rho_Q):=\diint\frac{\rho_Q(x)^*\rho_Q(y)}{|x-y|}dxdy,\ \text{Ex}[Q]:=\diint \frac{|Q(x,y)|^2}{|x-y|}dxdy.
\end{array}
\right.
\]
The function $\mathcal{E}_{\text{BDF}}^\nu$ is the BDF energy we will deal with in this paper. 

\begin{notation}Throughout this paper we write $\mathcal{P}^0_\eps Q \mathcal{P}^0_{\eps'}=Q^{\eps\,\eps'}$ where $\eps,\eps'\in\{+/-\}$.  For an operator $Q$ with integral kernel $Q(x,y)$, we define $R_Q$ by its integral kernel: $R_Q(x,y):=\tfrac{Q(x,y)}{|x-y|}$. There holds: $\nqq{Q}^2:=\text{Ex}[Q]=\ttr(R_Q^* Q)$. We write $\mathcal{C}$ the Hilbert space of densities with finite Coulomb energy:
\begin{equation}
\mathcal{C}:=\big\{\zeta\in\mathcal{S}'(\RR),\ \ncc{\zeta}^2:=4\pi \dint \dfrac{|\wh{\zeta}(k)|^2}{|k|^2}dk<+\infty\big\}.
\end{equation}
The squared norm $\ncc{\zeta}^2$ coincides with $\underset{\RR\times\RR}{\iint} \zeta^*(x)\zeta(y)\tfrac{dxdy}{|x-y|}$ whenever this last integral converges.
\end{notation}

A justification to study the BDF energy -- stated in \cite{mf} -- is the following. In the presence of an external charge density $\nu$ such that $D(\nu,\nu)<+\infty$ and that $\wh{\nu}$ continuous is in $B(0,\La)$, one can consider the corresponding minimization problem of $\mathbb{H}^\phi$ in $\mathfrak{H}_\La^L$. There also exists a minimizer with energy $E_{L}(\phi)$ and in the thermodynamic limit:
\[
\left\{
\begin{array}{l}
\underset{L\to+\infty}{\lim}(E_L(\phi)-E_L(0))=\underset{Q\in \mathcal{Q}_\La}{\inf}\mathcal{E}_{BDF}^{\nu}(Q),\\
\text{where\ }\mathcal{Q}_\La:=\{Q\in\mathfrak{S}_2(\hl),\,-\PP\le Q\le \PPP,Q^{++},Q^{--}\in\mathfrak{S}_1(\hl) \}.
\end{array}
\right.
\]

\begin{notation}
We recall that for each $1\le p\le+\infty$, $\mathfrak{S}_p(\hl)$ is the subspace of compact operators $A\in\mathcal{B}(\hl)$ with $\text{Tr} |A|^p<+\infty$. The case $p=1$ gives trace-class operators and $p=2$ gives Hilbert-Schmidt operators. We recall $Q$ is Hilbert-Schmidt if and only if its integral kernel is in $L^2(\hl\times\hl)$.
\end{notation}

Instead of minimizing over all states in $\mathcal{Q}_\La$, we may minimize over sector charge $\mathcal{Q}_\La(q),\,q\in\mathbf{R}:$
\begin{equation}\label{qla}
\mathcal{Q}_\La(q):=\{ Q\in \mathcal{Q}_\La, \text{Tr}(Q^{++}+Q^{--})=q\}.
\end{equation}
The number $q$ is interpreted as the number of electrons (if $q\in\mathbf{N}^*$) or the number of positrons (if $q\in\mathbf{Z}\backslash \mathbf{N}$). In the presence of an external field $\nu$, the energy function is then defined as
\begin{equation}\label{ebdf}
E_{\text{BDF}}^{\nu}(q):=\inf\big\{ \mathcal{E}_{\text{BDF}}^\nu(Q),\ Q\in \mathcal{Q}_\La(q)\big\}.
\end{equation}
In \cite{at}, Hainzl \emph{et al.} have shown that for any $q_0\in\mathbf{R}$, the problem $E_{\text{BDF}}^{\nu}(q_0)$ admits a minimizer as soon as there hold binding inequalities:
\begin{equation}\label{con-pact}
\forall\ q\in\mathbf{R}\backslash\{0\},\ E_{\text{BDF}}^{\nu}(q_0)<E_{\text{BDF}}^{\nu}(q_0-q)+E_{\text{BDF}}^{0}(q).
\end{equation}
A more difficult task is to check these inequalities hold for some $q_0$. In \cite{at}, by this method it is proved that for any $\nu\in L^1(\RR,\mathbf{R}_+)\cap \mathcal{C}$ and any integer $M$ such that $0\le M<\int\nu+1$, the problem $E_{\text{BDF}}^\nu(M)$ admits a minimizer (a so-called ground state) close to the limit $\alpha\to 0$ with $\La=\La_0>0$ kept \emph{fixed}.

In this paper we show there exists a minimizer for $E_{\text{BDF}}^0(1)$, provided $\alpha, \Lambda^{-1}$ and $\alpha\llo$ are sufficiently small. It is remarkable that the system of one electron in the Dirac sea can bind in the absence of any external field: this answers an open question stated in \cite{mf} (page 19). The presence of the electron induces the polarization of the Dirac sea: it is locally repelled in the neighbourhood of the particle. This fact is illustrated by the inequality $E_{\text{BDF}}^0(1)<m(\alpha)$ where $m(\alpha)$ is the infimum of the BDF energy among configurations where the Dirac sea, represented by $\PP$, is not polarized:
\[
m(\alpha)=\underset{\phi\in \text{Ker}\,\PP}{\inf}\EE(\ket{\phi}\bra{\phi})=\inf \sigma(|\D|).
\]

We are then interested in the non-relativistic limit $\alpha\to 0$ with $\alpha\llo$ kept fixed to a small value (it may not be $0$). The wave function $\psi$ of the real electron has a specific behaviour. There exists $c(\alpha,\La)>0$ with $c=O(\alpha^{-2}\llo^{-1})$ such that up to translation and up to scaling by $c>0$, the upper spinor of the wave function $\psi$ tends to a minimizer of the Choquard-Pekar energy $E_{\text{CP}}$ \cite{L}:
\begin{equation}\label{choqua}
E_{\text{CP}}:=\underset{\scriptstyle \phi\in H^1(\RR): \lVert \phi\rVert_{L^2}=1}{\inf}\left\{\mathcal{E}_{\text{CP}}(\phi):=\dint |\nabla \phi|^2dx-D(|\phi|^2,|\phi|^2)\right\}<0.
\end{equation}
More precisely the Choquard-Pekar energy $\mathcal{E}_{\text{CP}}$ of $\un{\psi}(x):=c^{3/2} \psi(c x)$ tends to $E_{\text{CP}}$. The link with a model of polaron is natural: the Dirac sea is a polarizable system and like a lattice of ions reacts to the presence of an electron. The smallness of $\alpha\llo$ corresponds to a small charge renormalisation. As explained in \cite[part 4]{Sc}, the \emph{physical} coupling constant $\alpha_{\text{phys}}$ is different from its "bare" value $\alpha$. More precisely in the reduced BDF model, where the exchange term is neglected, a minimizer of $\mathcal{E}^\nu_{\text{BDF}}$ with $\nu\ge 0$ radial (interpreted for instance as $\int\nu=Z$ protons) and $D(\nu,\nu)$ small enough has radial density $\rho_\g$ \cite{gs}, the potential induced by $\nu$ at infinity is not $\alpha Z\tfrac{1}{|x|}$ as it should be but rather $(\nu-\rho_\g)*\tfrac{1}{|\cdot|}(x)\underset{|x|\to+\infty}{\sim}\alpha_{\text{phys}}Z\tfrac{1}{|x|}$ where
\begin{equation}\label{charger}
\alpha_{\text{phys}}=\alpha Z_3,\ Z_3=\frac{1}{1+\alpha B_\Lambda^0(0)}\text{\ and\ }B_\Lambda^0(0)=\frac{2}{3\pi}\llo+O(1).
\end{equation}
The quantity $B_\Lambda(0)^0$ is the value at $k=0$ of the function defined in Notation \ref{blamb} below and $Z_3$ is the charge renormalization constant. If we assume the charge renormalization in the full model to be a perturbation of \eqref{charger}, fixing $0<\alpha\llo=L_0\ll 1$  corresponds to considering $0< 1-Z_3\ll 1$.

In this paper we have chosen the model of \cite{mf} with $\PP$ as reference state instead of that of \cite{ptf,Sc} with $P^0_-$ as reference state, but all the results proved here are also true in this last model with the same proofs.

The paper is organized as follows: in the next section we properly state the variational problem $E_{\text{BDF}}^0(1)$ and state the main theorems. Subsections \ref{banach} and \ref{someineq} are devoted to introduce the Banach spaces and the inequalities used throughout the paper. Theorem \ref{tlsup} gives an upper bound of $E_{\text{BDF}}^0(1)$ which is the BDF energy of a test function $\G$. This test function is defined by adapting the fixed point scheme in \cite{ptf}: the method is explained in Subsection \ref{explptf} and the needed estimates in Appendices \ref{needed} and \ref{estimfp}. Then Proposition \ref{HVZ} states that the binding inequalities at level 1 are true for $\EE$, as a consequence there exists a minimizer for $E_{\text{BDF}}^0(1)$. Theorem \ref{tlinf} gives a lower bound of $E_{\text{BDF}}^0(1)$ by computing the BDF energy of a minimizer. The two theorems and the proposition are proved in Section \ref{proofs}. At last we look at the nonrelativistic limit in Theorem \ref{nrlim}. Appendix \ref{operateurD} is devoted to prove estimates linked to the use of the operator $\D$.

\section{Description of the model and main results}\label{main}
We start with some definitions and notations. Our convention for the Fourier transform $\mathscr{F}$ is:
\[
\forall f\in \hl\cap L^1(\mathbf{R}^3,\mathbf{C}^4),\ \wh{f}(p):=\frac{1}{(2\pi)^{3/2}}\dint f(x)e^{-ix\cdot p}dx.
\]

\noindent In Fourier space $\D$ takes the following form
\begin{equation}
\wh{\D}(p)=\boldsymbol{\alpha}\cdot \om_p g_1(|p|)+g_0(|p|)\beta,\ \om_p=\dfrac{p}{|p|},
\end{equation}%(
where $g_0,g_1: [0,\La)\to \mathbf{R}_+$ are real and smooth functions satisfying%]
\begin{equation}\label{equivg}
x\le g_1(x)\le xg_0(x).
\end{equation}
It is possible to improve estimations of \cite{ls} in the regime $L:=\alpha\llo=O(1)$: we get estimates of the derivatives of $g_0,g_1$ by using their self-consistent equation (\emph{cf} Appendix A). We write $m(\alpha)$ for the bottom of $\sigma(|\D|)$:
\begin{equation}\label{defmalpha}
m(\alpha):=g_0(0)=\min(\sigma(|\D|)).
\end{equation}

We introduce the following notations concerning the Dirac operator:

\begin{notation}\label{gstar}
We write $\ed{p}:=\sqrt{g_0(p)^2+g_1(p)^2}=|\D(p)|$ and 

$\ee{p}:=\sqrt{1+|p|^2}=|D^0(p)|$.

We write $g_0$ (respectively $g_1)$ for both functions\newline
\noindent  $g_\star:x\in [0,\Lambda]\to g_\star(x)\in\mathbf{R}^+$ and $g_\star:p\in B(0,\Lambda)\to g_\star(|p|)\in\mathbf{R}^+$. The $(g_0)$'s are in $\mathcal{C^\infty}$ while $g_1\in \mathcal{C}^1(B(0,\Lambda))$ (\emph{cf} Appendix \ref{operateurD}).

At last we write
\[
\left\{
\begin{array}{l}
\mathbf{g_1}:p\in B(0,\Lambda)\to g_1(|p|)\om_p\in\RR\\
\mathbf{g}:p\in B(0,\Lambda)\to \begin{pmatrix}g_0(p)\\ \mathbf{g_1}(p) \end{pmatrix}\in\mathbf{R}^4.
\end{array}\right.
\]
\end{notation}
\begin{notation}\label{cun}
$C_1\ge 1$ denotes a constant satisfying $g_1(r)\le C_1 |r|$ and $|g_0|_{\infty}\le C_1$.
\end{notation}

\begin{notation}\label{blamb}
A recurrent function of this problem is 
\begin{equation}\label{blaeq}
B_\La(k):=\frac{1}{\pi^2|k|^2}\dint\limits_{\scriptstyle \lvert p=l+\tfrac{k}{2}\rvert,\lvert q= l-\tfrac{k}{2}\rvert\le \La} \frac{\ed{p}\ed{q}-\mathbf{g}(p)\cdot \mathbf{g}(q)}{\ed{p}\ed{q}(\ed{p}+\ed{q})}dl.
\end{equation}
If we replace $\ed{\cdot}$ by $E(\cdot)$ we get the function $B_\La^0$ of \cite{ptf,gs}. We define the function $\WW(k)$ by the formula
\begin{equation}\label{defww}
\WW(k):=\tfrac{\alpha B_\La(k)}{1+\alpha B_\La(k)}.
\end{equation}

\noindent In Appendix \ref{operateurD} it is shown that $B_\La(k)=O(\log(\La))$ and that for $L\ll 1$ there holds $B_\La(0)=\tfrac{2}{3\pi}\llo+O(L\llo+1)$.
\end{notation}

We consider then the $\PP$-trace ($\PP$ is defined in the introduction):
\begin{equation}\tr(Q):=\ttr(\PP Q\PP)+\ttr(\mathcal{P}^0_+Q\mathcal{P}^0_+),\ \mathcal{P}^0_+:=1-\PP.\end{equation}
As shown in\cite{ptf} we know the operators $Q^{--}=\PP Q\PP$ and $Q^{++}=\mathcal{P}^0_+Q\mathcal{P}^0_+$ are trace-class when $Q\in\mathfrak{S}_2(\hl)$ is a difference of two orthogonal projectors of the form $Q=P-\PP$. In this case:
\[
|Q|^2=Q^2=Q^{++}-Q^{--}.
\]

We introduce the set of $\PP$-trace class operators:
\[
\mathfrak{S}_1^{\PP}(\hl)=\big\{Q\in \mathfrak{S}_2(\hl):Q^{++},Q^{--}\in \mathfrak{S}_1(\hl)\big\}.
\]
The variational set $\mathcal{Q}_\La$ (\textit{cf} introduction) is a convex set of $\mathfrak{S}_1^{\PP}(\hl)$ and its extremal points are that of the form $Q=P-\PP$ where $P$ is an orthogonal projector.

The density of an operator $Q\in \mathcal{Q}_\La$ is $\rho_{Q}(x)=\ttr_{\CC}(Q(x,x))$. It is mathematically well defined since $Q$ is locally trace-class (thanks to the cut-off). The Fourier transform of $\rho_Q$ is:
\begin{equation}
\widehat{\rho_Q}(k):=\frac{1}{(2\pi)^{3/2}}\dint\limits_{|u+\tfrac{k}{2}|,|u-\tfrac{k}{2}|\le \La}\ttr_{\CC}(\hat{Q}(u+\tfrac{k}{2},u-\tfrac{k}{2}))du,
\end{equation}

In the absence of external field, the energy functional defined on $\mathcal{Q}_\La$ is 
\[\EE(Q)=\tr(\D Q)+\frac{\alpha}{2}\left(D(\rho_Q,\rho_Q)-\nqq{Q}^2\right).\]
The trace part is the kinetic energy while the two others are respectively the \emph{direct term} and the \emph{exchange term}. Moreover the following inequalities hold \cite{stab,ptf,mf} \begin{subequations}\label{eqin}
\begin{equation}\tr(\D Q)=\ttr(|\D|(Q^{++}-Q^{--}))\ge\ttr(|\D|Q^2),\end{equation}
\begin{equation}\label{eqinb}\iint\frac{|Q(x,y)|^2}{|x-y|}dxdy\le \frac{\pi}{2}\ttr(|\D|Q^2).\end{equation}\end{subequations}
 Inequality~\eqref{eqinb} is due to Kato's inequality\eqref{kato}. We assume that $\alpha<\tfrac{4}{\pi}$: in this case $\EE$ is bounded from below \cite{ptf}. 
 
 We study the variational problem $E_{\text{BDF}}^0(1)$. To ensure the existence of a minimizer for $E_{\text{BDF}}^0(1)$, it suffices to prove the following binding inequalities \cite{at}.

\begin{proposition}\label{HVZ}
There exist three constants $\alpha_0, L_0,\La_0>0$ such that

\noindent if $0<\alpha\le \alpha_0, 0<L\le L_0$ and $\La\ge \La_0$, then:
\begin{equation}
\forall\,q\in\mathbf{R}\backslash\{0,1\}:\ E_{\text{BDF}}^0(1)<E_{\text{BDF}}^0(1-q)+E_{\text{BDF}}^0(q).
\end{equation}
\end{proposition}

This Proposition comes as a corollary of the following Theorem. 
\begin{theorem}\label{tlsup}
There exist three constants $\alpha_0,L_0,\La_0>0$ such that

\noindent if $\alpha\le\alpha_0,L\le L_0,\La\ge\La_0$ then:
\begin{equation}
E_{\text{BDF}}^0(1)\le m(\alpha)+\frac{(\alpha \WW(0))^2m(\alpha)}{2g'_1(0)^2}E_{\text{CP}}+o((\alpha \WW(0))^2)<m(\alpha),
\end{equation}
where $E_{\text{CP}}$ is the Choquard-Pekar energy (see \eqref{choqua}).
\end{theorem}

\begin{remark}
For sufficiently small $\alpha\llo$ we have $g'_1(0)>\eps>0$. More generally all the results we need about $g_0$ and $g_1$ are proved in Appendix \ref{operateurD}.
\end{remark}

\begin{notation}
Throughout this paper we work in the regime
\begin{equation}\label{regi}
\alpha\ll 1,\ \La\gg 1,\ \alpha\llo=L\le \eps_0,
\end{equation}
so whenever we write $o(\cdot)$ and $O(\cdot)$ without specifying the limit it is understood that it holds in the regime \eqref{regi}. Moreover, $K$ denotes a constant which is independent of $\alpha$ and $\Lambda$. The inequality $a\apprle b$ means that $a\le Kb$ where $a$ and $b$ are positive real numbers.
\end{notation}

To understand what happens in Theorem \ref{tlsup} let us see what should be a minimizer of $E_{\text{BDF}}^0(1)$. We have the following lemma (proved in Section \ref{end?}, see Lemma \ref{1st_step})
\begin{lemma}\label{happens}
A minimizer $Q$ for $E_{\text{BDF}}^0(1)$ can be decomposed as $Q=\g +\ket{\psi}\bra{\psi}$ where $\g,\psi$ satisfy the self-consistent equations:
\begin{equation}\label{eqg}
\left\{
\begin{array}{rl}
\g+\PP&=\chi_{(-\infty,0)}(\mathcal{D}_Q),\ \mathcal{D}_Q:=\D +\alpha\Big(\rho_Q* |\cdot|^{-1}-\frac{Q(x,y)}{|x-y|}\Big),\\
\ket{\psi}\bra{\psi}&=\chi_{[0,\mu]}(\mathcal{D_Q}).
\end{array}
\right.
\end{equation}
The number $0<\mu<m(\alpha)$ can be chosen such that $\mathcal{D}_Q\psi=\mu\psi$. 
\end{lemma}
Thanks to Proposition 1 of \cite{at}, there only remains to prove $\chi_{[0,\mu]}(\mathcal{D_Q})$ has rank $1$: as $\g+\PP$ is a compact perturbation of $\PP$, its essential spectrum is the same and necessarily $0\le \mu<m(\alpha)$ and $\chi_{[0,\mu]}(\mathcal{D_Q})$ is the projection onto an eigenspace of $\mathcal{D}_Q$. It suffices to prove $\ns{2}{\g}=o(1)$ to get:
\[
\ttr\big(\chi_{[0,\mu]}(\mathcal{D_Q})\big)=\tr\big(\chi_{[0,\mu]}(\mathcal{D_Q})\big)=\tr(\g')-\tr(\g)=1.
\]

The strategy for Theorem \ref{tlsup} is to take a test function $\G$ which satisfies an equation similar to \eqref{eqg}. To this end let us first take $\phi'_1$ the \emph{unique} positive radial minimizer of the Choquard-Pekar energy (\textit{cf} Introduction) and consider $\phi_1:=\tfrac{P_{\hl}\phi_1'}{\nlp{2}{P_{\hl}\phi_1'}}$ where $P_{\hl}$ is the projector onto $\hl$. We consider the spinor: $\psi_1:=\begin{pmatrix}\phi_1\\ 0 \end{pmatrix}$ . For $\lambda^{-1}:=\tfrac{\alpha \WW(0)m(\alpha)}{g'_1(0)^2}$ we write
\begin{equation}\label{testn}
\psi_\la:=\la^{-3/2}\psi_1(\la^{-1}(\cdot)),\ N=N_\la:=\ket{\psi_\la}\bra{\psi_\la}\text{\ and\ }n_\la:=|\psi_\la|^2=\rho_N. 
\end{equation}
It is possible to adapt the fixed point method of \cite{ptf} to define $\g$ as the solution to 
\begin{equation}\label{testg}
\gamma=\chi_{(-\infty,0)}\Big(\D+\alpha((\rho_{\g}+n)* |\cdot|^{-1}-R[\g+N])\Big)-\PP,\ \ \ \ \ \ \ \ \ \ \ \ 
\end{equation}
provided $\alpha$ and $\alpha\llo$ are small enough. In fact this paper \cite{ptf} treats the case of $D^0$ but in Appendix \ref{estptf} it is shown that replacing it by $\D$ is harmless (\emph{cf} Lemmas \ref{fpq} and \ref{fpqu}).

We chose the test function $\G$ defined by the formulae
\begin{equation}\label{test2}
\G:=\g+N',\ \pi=\g+\PP,\text{\ and\ } N'=\frac{\ket{(1-\pi)\psi_\la}\bra{(1-\pi)\psi_\la}}{1-\lVert\pi\lpsi\rVert_{L^2}^2}.
\end{equation}

We then compute compute $\EE(\G)$ using that an electron does not see its own field (that is here $D(|\psi|^2,|\psi|^2)-\text{Ex}\big[\ket{\psi}\bra{\psi}\big]=0$).
\begin{lemma}\label{llsup}
Let $\G$ be as above \eqref{testg}, \eqref{test2}. Then the following estimate holds:
\begin{equation}\label{testfun}
\EE(\G)=m(\alpha)+\frac{\alpha \WW(0)}{2\la}E_{\text{CP}}+o\left(\frac{\alpha \WW(0)}{\la}\right).
\end{equation}
More precisely, writing $I=\ncc{\rho_\G}^2-\ncc{\rho_{N'}}^2$ and $J=\text{Ex}[\G]-\ncc{\rho_{N'}}^2$ we have
\[\begin{array}{|rl}
\tr(\D N')&=m(\alpha)+\frac{g'_1(0)^2}{2\la^2 m}\dint |\nabla \psi_1|^2 dx+o(\la^{-2}),\\
\tr(D \g)&=\frac{\alpha(\WW(0)-\WW(0)^2)}{2\la}D(n_1,n_1)+o\left(\frac{\alpha \WW(0)}{\la}\right),\\
\dfrac{\alpha}{2}I&=-\frac{\alpha (2\WW(0)-\WW(0)^2)}{2\la}D(n_1,n_1)+o\left(\frac{\alpha \WW(0)}{\la}\right),\\
\alpha J&=o\left(\frac{\alpha \WW(0)}{\la}\right).
\end{array}\]
\end{lemma}
Lemma \ref{llsup} is proved in Section \ref{pllsup}. Theorem \ref{tlsup} is an obvious corollary.

At this point we know there exists a minimizer $\g'=\g +\ket{\psi}\bra{\psi}$ for $E_{\text{BDF}}^0(1)$ and it satisfies Eq.~\eqref{eqg}. The computation of its energy in terms of $\psi$ gives a lower bound of $E_{\text{BDF}}^0(1)$ of the same form as the right hand side of \eqref{testfun}.
\begin{theorem}\label{tlinf}
There exist three constants $\alpha_1,L_1,\La_1>0$ such that for $\alpha\le\alpha_1$, 

\noindent$L\le L_1$, $\La\ge \La_1$, there holds
\begin{equation}
E_{\text{BDF}}^0(1)=m(\alpha)+\frac{(\alpha \WW(0))^2m(\alpha)}{2(g'_1(0))^2}E_{\text{CP}}+o\left((\alpha \WW(0))^2\right).
\end{equation}
\end{theorem}

\begin{theorem}\label{nrlim}
Writing $C_0^2:=\frac{2g'_1(0)^2}{(\alpha \WW(0))^2m(\alpha)}$ in the regime \eqref{regi} we have:
\begin{equation}\label{fin}
\liminf\limits_{\alpha,\La^{-1}\to 0} C_0^2(E_{\text{BDF}}^0(1)-m(\alpha))=\limsup\limits_{\alpha,\La^{-1}\to 0} C_0^2(E_{\text{BDF}}^0(1)-m(\alpha))=E_{\text{CP}}.
\end{equation}
Assume $Q$ is a minimizer for $E_{\text{BDF}}^0(1)$: as in \eqref{eqg} we can write: $Q=\g+\ket{\psi}\bra{\psi}$. In the limit $\alpha\to 0$ where $\alpha\llo=L'$ is kept \emph{fixed}  and for $L'$ small enough the following holds:

Up to translation, the upper spinor $\un{\ph}\in H^1(\RR,\mathbf{C}^2)$ of $\un{\psi}(x):=c^{3/2}\psi(cx)$ tends to a minimizer of the Choquar-Pekar energy $E_{\text{CP}}$.
\end{theorem}

%\begin{remark}$\alpha\llo=L'$ is fixed and
%If $L=\alpha \llo\le \min(L_0,L_1)$ is fixed then $\lim_{\alpha\to 0}\La^{-1} = 0$ so \eqref{fin} holds with $\alpha\to 0$.
%\end{remark}

\begin{remark}
This paper uses heavily estimates and proofs of \cite{ptf}. For convenience Lemma \ref{nonlin} is not fully proved: it is an adaptation of \cite{ptf}, the whole proof is in the thesis \cite{these} of the author.
\end{remark}

\section{Preliminary results}

\subsection{Banach spaces}\label{banach}
%(
In this paper several Banach spaces are used.

As usual $\nlp{p}{\cdot}$ and $\nso{s}{\cdot}$ for $p\in[1,+\infty)$ and $s\in\mathbf{R}_+$ are the usual norms of $L^p$ and Sobolev functions. Moreover $\ns{p}{\cdot}$ is the norm of the space of Schatten-class operators $\mathfrak{S}_p(\hl)$ and $\nb{\cdot}$ is the usual norm of bounded linear operators in $\mathcal{B}(\hl)$. The norms $\ncc{\cdot}$ and $\nqq{\cdot}$ are defined in the introduction.

\noindent A large part of the paper is devoted to estimate Sobolev norms of test functions $Q$ and among them the norm
\begin{equation}
\nqkin{Q}^2:=\ttr(|\D| |Q|^2)
\end{equation}
is linked to the kinetic energy of $Q$. 

In \cite{ptf} Hainzl \emph{et al.} introduce the following norms for $(Q,\rho)\in \mathfrak{S}_2(\hl)\times \mathcal{C}\cap L^2$:
\begin{equation}
\left\{
\begin{array}{rl}
\nq{Q}^2&:=\diint \ed{p-q}^2\ed{p+q}|\wh{Q}(p,q)|^2dpdq,\\
\nc{\rho}^2&:=\dint\frac{\ed{k}^2|\wh{\rho}(k)|^2}{|k|^2}dk\apprle \ncc{\rho}^2+\nlp{2}{\rho}^2.
\end{array}
\right.
\end{equation}
Strictly speaking, the authors use $E(\cdot)$ instead of $\ed{\cdot}$. However thanks to \eqref{equivg} and \eqref{defmalpha} these norms are equivalent:
\[
\exists K>0,\ \forall\,p\in B(0,\La),\ \frac{1}{K}E(p)\le \ed{p}\le KE(p).
\]
Moreover we write for an operator $R(x,y)$:
\begin{equation}
\nqr{R}^2:=\iint \tfrac{\ed{p-q}^2}{\ed{p+q}}|\wh{R}(p,q)|^2dpdq.
\end{equation}
As in \cite{ptf}, we will estimate the above norm of $R_Q(x,y)=\tfrac{Q(x,y)}{|x-y|}$.

Unfortunately this is not sufficient and intermediate norms between $\nqkin{\cdot}$ and $\nq{\cdot}$ (respectively $\ncc{\cdot}$ and $\nc{\cdot}$) are necessary:
\begin{equation}
\left\{
\begin{array}{rl}
\nqu{Q}^2&:=\diint \ed{p-q}\ed{p+q}|\wh{Q}(p,q)|^2dpdq,\\
\nqo{Q}^2&:=\diint \ed{p+q}|\wh{Q}(p,q)|^2dpdq,\\
\ncu{\rho}^2&:=\dint\frac{\ed{k}|\wh{\rho}(k)|^2}{|k|^2}dk.
\end{array}
\right.
\end{equation}
%(
The numbers $0$ and $1$ refer to the exponent of $\ed{p-q}$ and $\ed{k}$.

We also introduce:
\begin{equation}
\nhi{Q}^2:=\diint \max\big\{\ed{p}, \ed{p-q}^2,\ed{p-q}\ed{p+q}\big\}|\wh{Q}(p,q)|^2dpdq.
\end{equation}
For any operator $Q\in\mathfrak{S}_2$ we have:
\begin{equation}
\sqrt{\frac{2}{\pi}}\nqq{Q}\le \nqkin{Q}\le \nhi{Q}\le \nq{Q}.
\end{equation}

For some function $f:\mathbf{R}^3\to [1,+\infty)$, we write:%]
\[
\begin{array}{ll}
\nqf{Q}^2:=\dint f(p-q)\ed{p+q}|\wh{Q}(p,q)|^2dpdq,&\ncf{\rho}^2:=\dint\frac{f(k)}{|k|^2}|\wh{\rho}(k)|^2dk.
\end{array}
\]

\subsection{Some inequalities}\label{someineq}

Let us recall Hardy's and Kato's inequalities we will use throughout this paper.
For $\phi\in L^2(\RR)$, the following inequalities hold:
\begin{subequations}
\begin{equation}\label{hardy}
\dint\frac{|\phi(x)|^2}{|x|^2}dx\le 4\psh{|\nabla|^2 \phi}{\phi},
\end{equation}
\begin{equation}\label{kato}
\dint\frac{|\phi(x)|^2}{|x|}dx\le \frac{\pi}{2}\psh{|\nabla| \phi}{\phi},
\end{equation}
\end{subequations}

Another recurrent inequality is Kato-Seiler-Simon's inequality (K.-S.-S.) \cite{Sim}: 

\noindent for any $f,g\in \mathcal{B}(\RR,\mathbf{\CC})$ (Borelian functions), we have:
\begin{equation}\label{K.-S.-S.}
\ns{p}{f(x)g(i\nabla)}\le (2\pi)^{-\tfrac{3}{p}}\nlp{p}{f}\nlp{p}{g},\,2\le p<\infty.
\end{equation}

We use the following Sobolev inequalities in this paper (\emph{cf} \cite{fouran} Theorem 1.38 p.29): for suitable $f$ ($f\in H^1(\mathbf{R}^3)$ for instance)
\begin{equation}\label{sobin}
\nlp{6}{f}\apprle \nlp{2}{\nabla f},\ \nlp{4}{f}\apprle \big\lVert\,|\nabla|^{3/4} f\big\rVert_{L^2},\ \nlp{3}{f}\apprle \big\lVert\,|\nabla|^{1/2} f\big\rVert_{L^2}.
\end{equation}

An immediate result of \eqref{K.-S.-S.} ($p=6$) and \eqref{sobin} ($p=3$) is the following Lemma.
\begin{lemma}\label{bs6}
Let $\rho\in\mathcal{C}$ and $\ph_{\rho}:=\rho*|\cdot|^{-1}$. For any $t>1/2$ there exists $K_t>0$ such that
\[
\ns{6}{\ph_{\rho}|\D|^{-t}}\le K_t\ncc{\rho}.
\]
Moreover we have:
\[
\begin{array}{ll}
\ns{6}{\ph_{\rho}|\D|^{-\tfrac{1}{2}}}\apprle (\llo)^{\tfrac{1}{6}}\ncc{\rho},& \nb{\ph_{\rho} |\nabla|^{-\tfrac{1}{2}}}\apprle \ncc{\rho}
\end{array}
\]
\end{lemma}
\begin{remark}
The notation $\ph_\rho$ is used throughout the paper.
\end{remark}

\noindent Let us consider $R=R_Q$ with $Q\in\mathcal{Q}_\La$. The Lemma 8 of \cite{ptf} states that:
\begin{equation}\label{lem8}
\nqr{R_Q}\apprle \nq{Q}.
\end{equation}
The following Lemma generalizes this result:
\begin{lemma}\label{rrq} Let $t\ge 0$. Then we have:
\begin{subequations}
\begin{equation}\label{a1}
\ns{2}{|\nabla|^{-1/2} R_Q}\apprle \nqq{Q},
\end{equation}
\begin{equation}\label{a2}
\diint \frac{\ed{p-q}^t}{\ed{q}^2}|\wh{R}(p,q)|^2dpdq\apprle \diint \ed{p-q}^t\ed{p+q}|\wh{Q}(p,q)|^2dpdq,
\end{equation}
\begin{equation}\label{a3}
\diint \frac{|\wh{R}(p,q)|^2}{\ed{q}}dpdq\apprle \diint \ed{p-q}\ed{p+q}|\wh{Q}(p,q)|^2dpdq.
\end{equation}
\end{subequations}
\end{lemma}
\noindent \textbf{Proof:} Ineq.~\eqref{a3} is a consequence of \eqref{lem8} for $\ed{q}^{-1}\apprle \tfrac{\ed{p-q}}{\ed{p+q}}$. 
 Ineq.~\eqref{a2} can be proved by adapting the proof of Lemma $8$.\cite{ptf} (see Lemma \ref{change8}). This gives:
\[
\diint \frac{\ed{p-q}^t}{\ed{q}^2}|\wh{R}(p,q)|^2dpdq\le 8\diint \ed{2v}^t\ed{2\ell}w(\ell,v)|\wh{Q}(\ell+v,\ell-v)|^2d\ell dv,
\]
where $w(\ell,v)$ is a weight lesser than
\[
\ed{2\ell}(2\pi^2)^{-2}\diint dud\ell'\big\{\ed{u-v}^2\ed{2\ell'}^{1+1}|\ell-u|^2|\ell'-u|^2\big\}^{-1}\apprle 1.
\]
 Ineq.~\eqref{a1} is proved as follows: up to a constant the operator $|\nabla|^{-1}$ acts in Direct space as a convolution by $\tfrac{1}{|\cdot|^2}$ (\emph{cf} \cite{LL}, p.130).

The operator $R_Q^*\tfrac{1}{|\nabla|} R_Q$ is nonnegative and by Cauchy-Schwartz inequality:
\[
\begin{array}{rl}
\ttr\big\{R_Q^*\tfrac{1}{|\nabla|} R_Q\big\}&\le \underset{(\mathbf{R}^3)^3}{\diiint} \dfrac{|Q(x,y)|}{|x-y|}\dfrac{dxdydz}{|y-z|^2}\dfrac{|Q(z,x)|}{|z-x|}\\
                   &\le \bigg\{\underset{(\mathbf{R}^3)^3}{\diiint} |Q(x,y)|^2\dfrac{dxdydz}{|y-z|^2|z-x|^2}\bigg\}^2\\
                   &\apprle \diint \frac{|Q(x,y)|^2}{|x-y|}dxdy. 
\end{array}
\]
\hfill{\footnotesize$\Box$}
\begin{lemma}\label{squareroot}
There exist $0<\eps<1$ and $K_0>0$ such that, for all $(Q,\rho)\in \text{Ex}\times\mathcal{C}$, if $\alpha(\nqq{Q}+\ncc{\rho})<\eps$, then
\begin{equation}
|\D|\big(1-\alpha K_0(\nqq{Q}+\ncc{\rho}) \big)\le|\D+\alpha (\ph_\rho-R_Q)|\le |\D|\big(1+\alpha K_0(\nqq{Q}+\ncc{\rho}) \big).
\end{equation}
\end{lemma}
\noindent \textbf{Proof:} We have 
\[
\nb{R_Q |\D|^{-1}}\le \ns{2}{R_Q|\D|^{-1}}\apprle \nqq{Q}\text{\ and\ }\nb{\ph_\rho |\D|^{-1}}\apprle \ncc{\rho}.
\]

As shown in \cite{ptf}, it suffices to take the square root of
\[
|\D|\big(1-2\alpha K(\nqq{Q}+\ncc{\rho})\big)\le |\D+\alpha (\ph_\rho-R_Q)|^2\le |\D|^2\big(1+\alpha K(\nqq{Q}+\ncc{\rho})\big)^2.
\]
\hfill{\footnotesize$\Box$}

\subsection{The fixed point method}\label{explptf}

In \cite{ptf} the authors prove the existence of a global minimizer of $\mathcal{E}_{\text{BDF}}^\nu$ under some assumptions on $\alpha,\La,\ncc{\nu}$. The authors show there exists a solution to the self-consistent equation that should satisfy a minimizer $Q_0$ of $E_{\text{BDF}}^\nu$ (when $P^0_-$ is taken as reference state). This equation is:
\[
Q_0+P^0_-=\chi_{(-\infty,0)}\big(D^0+\alpha ((\rho_{Q_0}-\nu)*\tfrac{1}{|\cdot|})-R_{Q_0}\big).
\]
To this end a fixed-point scheme based on this equation is used: let us adapt this proof to our problem.

As shown in \cite{ptf} we can use the Cauchy's expansion to write (at least formally)
\begin{subequations}
\begin{equation}\label{eqvac}
\widetilde{Q}=\chi_{(-\infty,0)}(\D +\alpha (\ph_Q-R_Q))-\chi_{(-\infty,0)}(\D)=\ssum_{k=1}^\infty \alpha^k Q_{k},
\end{equation}
\begin{equation}
Q_k=-\frac{1}{2\pi}\dint_{-\infty}^{+\infty}\dd\eta \frac{1}{\D+i\eta}\Big((R_Q-\ph_Q)\frac{1}{\D+i\eta}\Big)^k.
\end{equation}
\end{subequations}
We also expand $(R-\ph)^k$, $Q_k:=\sum_{j=0}^k Q_{j,j-k}$: the function $Q_{j,j-k}(\cdot,\cdot)$ is polynomial of degree $j$ in $R_Q$ and polynomial of degree $(j-k)$ in $\ph_Q$. Thanks to Lemmas \ref{bs6} and \ref{rrq} we know that each integral converges at least in $\mathfrak{S}_6(\hl)$. If we take the density of each $Q_k$, we also obtain a (formal) expansion of $\rho[\widetilde{Q}]$:
\begin{equation}
\rho[\widetilde{Q}]=\ssum_{k=1}^{+\infty}\alpha^k \rho_k=\ssum_{k=1}^{+\infty}\ssum_{j=0}^{k}\alpha^{k}\rho_{j,j-k}.
\end{equation}

In \cite{ptf} it is proved that provided $\alpha (\nq{Q}+\nc{\rho_Q})$ is small enough, those sums converge in $\mathcal{Q}$ for $\widetilde{Q}$ and in $\mathfrak{C}$ for $\rho[\widetilde{Q}]$. In fact the authors show:
\begin{proposition}\label{continuous}
For any $k\in\mathbf{N}^*$ and any $0\le j\le k$, the function
\[
F_{k,j}:
\begin{array}{cll}
\mathcal{Q}\times\mathfrak{C}&\to & \mathcal{Q}\times\mathfrak{C}\\
 (Q,\rho)&\to&\big(Q_{j,k-j}[Q,\rho], \rho_{j,k-j}[Q,\rho]\big) 
\end{array}
\]
is a \emph{continuous} polynomial operator (with estimates of the norm precised in Lemmas \ref{lin} and \ref{nonlin} in Appendix \ref{needed}).
\end{proposition}
We prove a similar result in the cited Lemmas.

It is necessary to precise the particular form of $\rho_{0,1}[\rho]$. A direct computation in Fourier space gives the following formula \cite{ptf}.
\begin{lemma}\label{computrho01}
For $\rho\in\mathcal{C}$ we have:
\[
\wh{\rho}_{0,1}(\rho;k)=-B_\La(k)\wh{\rho}(k)\in\mathcal{C}.
\]
If $\rho$ is in $\mathfrak{C}$ (respectively $\mathfrak{C}_1$) then so is $\rho_{0,1}[\rho]$. 
\end{lemma}
\noindent The last statement follows from the fact that $|B_\La(k)|\apprle \llo$, proved in Appendix \ref{operateurD}.

Let us describe a fixed-point scheme adapted to our problem in the spirit of \cite{ptf}. Given the projector $N$ that corresponds to the "real" electrons and $n=\rho_N$ its density, we try to define the dressed vacuum $Q$ surrounding it. We seek a solution to
\begin{equation}\label{cauchy0}
Q+\PP=\chi_{(-\infty,0)}\big(\D+\alpha (\ph_{Q+N}-R(Q+N))\big).
\end{equation}

For convenience we write $\rho'=\rho'_\g:=\rho+n$, $Q'=Q+N$, $\ph'_Q=\ph_{Q'}$; Eq.~\eqref{cauchy0} can be rewritten:
\begin{equation}\label{cauchy1}
F_Q(Q',\rho'):=\chi_{(-\infty,0)}(\D +\alpha (\ph'_Q-R_Q'))-\chi_{(-\infty,0)}(\D)+N=N+\ssum_{k=1}^\infty \alpha^k Q_{k}(Q',\rho').
\end{equation}
Taking the density $\rho$ of both sides and using Lemma \ref{computrho01} we get $\rho_{Q'}=F_\rho(Q',\rho')$ with:
\begin{equation}\label{cauchy2}
\wh{F_\rho}(k):=\frac{1}{1+\alpha B_\La(k)}\Big(\wh{n}(k)+\alpha \wh{\rho}_{1,0}(Q';k)+\ssum_{\ell\ge 2}\alpha^\ell\wh{\rho_\ell}(Q',\rho';k)\Big).
\end{equation}

We must precise the domain of the function 
\begin{equation}
F:=F_Q\times F_\rho.
\end{equation}

Following \cite{ptf} we first consider the Banach space $\mathcal{X}=\mathcal{Q}\times \mathfrak{C}$ with the norm
\[
\lVert(Q,\rho)\rVert_{\mathcal{X}}= 2C_1^{3/2}(2\sqrt{2}\nc{\rho}+C_R\sqrt{2}\nq{Q}),
\]
where $C_R>0$ is defined in \cite{ptf} and $C_1\ge 1$ is defined in Notation \ref{cun}.

\begin{lemma}\label{fpq}
There exist $R_\La,\eps_1,\eps_2>0$ such that if $\sqrt{L\alpha}\le \eps_1, \alpha \nx{(N,n)}\le \eps_2$ then $B_{\mathcal{X}}(0,R_\La)$ is $F$-invariant. The number $R_\La$ is $O(\sqrt{\llo})$. Moreover in this ball $F$ is Lipschitz with constant $\nu_0=O(\sqrt{L\alpha})$. In other words the fixed point theorem can be applied to $F$ on $B_{\mathcal{X}}(0,R_\La)$.
\end{lemma}
This lemma and the next one are proved in Appendix \ref{needed}.
%(
\begin{remark}\label{choicebanach}
As explained and proved in Appendix \ref{needed}, by adapting the estimates of \cite{ptf} we realize that another choice of norms for $F$ is possible and so another choice of Banach space on which applying the Banach fixed point theorem. Indeed let us take a radial function $f:\RR\to [1,+\infty)$: as long as there exists a constant $C(f)\ge 1$ such that
%]
\begin{equation}\label{conditions}\tag{\text{Cnd}}
\sqrt{f(p-q)}\le C(f)(\sqrt{f(p-p_1)}+\sqrt{f(p_1-q)}),
\end{equation}
we can apply the fixed point theorem with the norms
\[
\nqf{Q}^2:=\diint f(p-q)\ed{p+q}|\wh{Q}(p,q)|^2dpdq,\ \ncf{\rho}^2:=\dint\frac{f(k)|\wh{\rho}(k)|^2}{|k|^2}dk.
\]

Here we are interested in the case $f(p-q)=\ed{p-q}$ and $f(p-q)=1$.
\end{remark}

Let $\mathcal{X}_f\subset\mathfrak{S}_2(\hl)\times \mathcal{C}$ be the Banach space with norm 

\noindent $\nxf{(Q,\rho)}:=K(f)(\nqu{Q}+\nxu{\rho})$, for some $K(f)>0$ depending on $f$ (Appendix \ref{needed}).
\begin{lemma}\label{fpqu}
There exist $R'_\La,\eps_1',\eps'_2>0$ such that if $L\le \eps_1', \alpha \nxu{(N,n)}\le \eps_2'$ then $B_{\mathcal{X}_f}(0,R_\La')$ is $F$-invariant. The number $R_\La'$ is $O(\sqrt{\llo})$. Moreover in this ball $F$ is Lipschitz with constant $\nu_0'=O(\sqrt{L\alpha})$.
\end{lemma}

\section{Proofs}\label{proofs}

We will use the following Lemma, proved in Appendix \ref{estptf} (Subsection \ref{estimfp}).

\begin{lemma}\label{florilege_estimations}
Let $\psi_\la,\g,\rho_\g$ defined in \eqref{testn} and \eqref{testg}. Then the following estimates hold:
\begin{equation}
\begin{array}{rlrl rl}
\nq{\g}&\apprle \alpha                         &\nhi{\g}&\apprle L\alpha, &\ns{2}{\g}&\apprle \alpha\sqrt{L\alpha},\\
 \nc{\rho_\g}&\apprle L\sqrt{L\alpha},&\ncc{\rho_\g}&\apprle L\sqrt{L\alpha}.& &
\end{array}
\end{equation}
Moreover:
\begin{equation}
\nlp{2}{\g |\D| \psi_\la}+\nlp{2}{\g\psi_\la}\apprle \alpha\sqrt{L\alpha}\text{\ and\ }  \big\lVert\,\big[|\D|, \g\big] \big\rVert_{\mathfrak{S}_2}\apprle L\alpha.
\end{equation}
\end{lemma}

\subsection{Proof of Lemma \ref{llsup}}\label{pllsup}

We recall $N$ and $n$ are defined in \eqref{testn}.

\begin{notation}
For convenience we write 
\[
\phi_\la:=\dfrac{(1-\pi)\psi_\la}{\nlp{2}{(1-\pi)\psi_\la}}=\dfrac{(1-\pi)\psi_\la}{\sqrt{1-\nlp{2}{\pi\lpsi}^2}}.
\]
So we have $N'=\ket{\phi_\la}\bra{\phi_\la}$. Moreover we write
\begin{equation}
\tau:=\alpha \WW(0).
\end{equation}
\end{notation}

\begin{remark}\label{p0m}
Here $\la^{-1}$ and $\tau$ are of the same order $L\alpha$. A direct calculation shows that $\nlp{2}{\PP|\D| \psi_\la}=O(\la^{-1})$ and $\nlp{2}{|\D|\lpsi}=O(1)$. We will often use 
\begin{equation}
\nlp{2}{\pi\lpsi}\le \nlp{2}{\g\lpsi}+\nlp{2}{\PP\lpsi}\apprle (o(\tau)+\la^{-1}).
\end{equation}
\end{remark}

\noindent\textbf{1. Estimation of $J$}

Lemma  \ref{florilege_estimations} gives $\nqq{\g}^2\apprle \nhi{\g}^2=O(\tau^2)$. By Cauchy-Schwarz inequality and  Ineq.~\eqref{hardy}: for any $G=\ket{f}\bra{g}$ with $f,g\in H^1$
\[
|\ttr\big(G^*R_\g\big)|\le \min(\nqq{\g}\nqq{G},2\ns{2}{\g}\big\lVert\nabla f\big\rVert_{L^2}\nlp{2}{g}).
\]
Now thanks to  Ineq.~\eqref{kato} and Lemma \ref{florilege_estimations}:
\[
\begin{array}{rl}
\big\lVert\,|\pi \psi_\la|^2\big\rVert_{\mathcal{C}}^2&\apprle \nlp{2}{\pi\lpsi}^2\psh{|\D|\pi\lpsi}{\pi\lpsi}\\
\psh{|\D|\pi\lpsi}{\pi\lpsi}&\le 2\nlp{2}{\lpsi}\Big(\big\lVert\,\big[|\D|,\g\big]\,\big\rVert_{\mathfrak{S}_2}+\big\lVert\g |\D|\lpsi\big\rVert_{L^2}\Big)=O((L\alpha)^2).
\end{array} 
\]

Similarly we have:
\[
\big|D\big(|\lpsi|^2,|\pi\lpsi|^2\big)\big|\apprle \nlp{2}{\pi\lpsi}^2 \psh{|\nabla|\lpsi}{\lpsi}\apprle \tfrac{(\tau+\la^{-1})}{\la},
\]
\[ \text{and}:\ 
|\ttr(R_\g^* N')|\le 2\ns{2}{\g}\nlp{2}{\nabla \lpsi}\nlp{2}{\lpsi}\apprle \tau \la^{-1}.
\]

\noindent Thus $J=O(\tau^2+\la^{-2})=O((L\alpha)^2)$.

\noindent\textbf{2. Estimation of $I$}

According to the self-consistent equation satisfied by $\rho_\g$, we write
\begin{equation}
\wh{\rho}(\g;k)=-\WW(k)\wh{n}(k)+(1-\WW(k))\wh{\rho}_{1,0}(\g;k)+(1-\WW(k))\ssum_{\ell=2}^\infty \alpha^\ell \wh{\rho}_\ell(\g;k)
\end{equation}
where we recall that $\WW(p)=\tfrac{\alpha B_\La(p)}{1+\alpha B_\La(p)}$. We write $\rho_\ell:=\rho_\ell(\g)$ and $\sum:=\sum_{\ell=2}^{+\infty}\alpha^{\ell}\rho_\ell$ for short. There holds:
\[
\begin{array}{l}
D(\rho_\g,\rho_\g) =4\pi\dint_k \dfrac{dk}{|k|^2}\bigg(\WW(k)^2|\ww{n}(k)|^2+(1-\WW(k))^2|\alpha \wh{\rho}_{1,0}(k)|^2+(1-\WW(k))^2\big|\wh{\sum}\big|^2\\
 \ \ \ \ \ \ +2\rr\left(\WW(k)(1-\WW(k))\overline{\ww{n}(k)}\left(\alpha \wh{\rho}_{1,0}(k)+\wh{\sum}\right)+(1-\WW(k))^2\overline{\alpha \wh{\rho}_{1,0}(k)}\wh{\sum}\right)\bigg).
 \end{array}
\]
By Cauchy-Schwarz inequality it suffices to study $\int \tfrac{|\wh{\rho}(k)|^2}{|k|^2}dk$ for $\rho\in\{n,\rho_{1,0},\sum\}$. We recall $\nlp{\infty}{\WW}\apprle L<2^{-1}$ for sufficiently small $L$.

\begin{lemma}\label{chchch}
Let $i\in\{1,2\}$, then there holds:
\begin{equation}
4\pi\dint_p\WW(p)^i\dfrac{|\wh{n_\la}(p)|^2}{|p|^2}dp=\WW(0)^i\frac{D(n_1,n_1)}{\la}+\underset{\la\to\infty}{o}(L^i\la^{-1}).
\end{equation}
Moreover:
\begin{equation}
\begin{array}{rl}
\alpha\ncc{\rho_{1,0}(\g)}&\apprle \sqrt{L\alpha}\nhi{\g}\apprle (L\alpha)^{-3/2},\\
\ncc{\sum}&\apprle \alpha^2.
\end{array}
\end{equation}
\end{lemma}

\noindent Before proving this Lemma, we show the estimation of $I$. First there holds:
\[
\ncc{\rho_\g}^2=\frac{\WW(0)^2}{\la}D(n_1,n_1)+\underset{\la\to\infty}{o}\left(\frac{L}{\la}\right).
\]

Then $|\phi_\la|^2(x)=\tfrac{1}{1-\nlp{2}{\pi\lpsi}^2}(|\lpsi(x)|^2+|\pi\lpsi(x)|^2-2\mathfrak{R}\,\lpsi^*(x)(\pi\lpsi)(x))$. By Cauchy-Schwarz and Kato inequalities the two last terms are $O(L(L\alpha)^2)$. In fact:
\[
\begin{array}{rl}
\big\lVert\,|\pi \lpsi|^2\,\big\rVert^2_{\mathcal{C}}&\apprle \nlp{2}{\pi\lpsi}^2\psh{|\nabla| \pi\lpsi}{\pi\lpsi}\apprle (L\alpha)^4\\
\big\lVert\,|\pi\lpsi| |\lpsi|\,\big\rVert^2_{\mathcal{C}}&\le 2\nlp{2}{\pi\lpsi}^2\nlp{2}{\nabla\lpsi}\nlp{2}{\lpsi}\apprle (L\alpha)^3,
\end{array}
\]
\noindent so $\big|D\big(\rho_\g,|\pi\lpsi|^2-2\mathfrak{R}\{\lpsi^*(\pi\lpsi)\}\big)\big|\apprle L\sqrt{L\alpha}(L\alpha)^{3/2}$.

Then $D(\rho_\g,n_\la)=-4\pi\int \WW(k)|\wh{n_\la}(k)|^2\tfrac{dk}{|k|^2}+O\big\{(\alpha\ncc{\rho_{1,0}}+\ncc{\sum})\ncc{n_\la}\big\}.$

In the same way:
\[
D(\rho_\g,|\phi_\la|^2)=-\WW(0)D(n_\la,n_\la)+\underset{\la\to\infty}{o}(\tfrac{L}{\la}).
\]
Since $\tfrac{1}{1-\nlp{2}{\pi\lpsi}^2}=1+O((\tau+\la^{-1})^2)$, we finally obtain:
\begin{equation}
I=-\frac{2\WW(0)+\WW(0)^2}{\la}D(n_1,n_1)+\underset{\la\to\infty}{o}\left(\frac{L}{\la}\right)
\end{equation}

\noindent\textbf{Proof of Lemma} \ref{chchch}.  We use Proposition \ref{z(x)} (Appendix \ref{operateurD}). In the regime \eqref{regi} and for $\eps=\tfrac{1}{6}$, in a neighbourhood $B(0,r_\eps)$ of $0$ \emph{independent} of $\alpha,\La$ we have: 
\begin{equation}
\forall\,k\in\ B(0,r_\eps)\backslash\{0\},\ \frac{|B_\La(|k|)-B_\La(0)|}{|k|}\apprle (\La^{-1}+|k|^{1/2})=:z(|k|).
\end{equation}
Then
\[
\dint_k\frac{\WW(k)^2|\wh{n_\la}(k)|^2}{|k|^2}dk=\dfrac{1}{\la}\dint_k\frac{\WW(\tfrac{k}{\la})^2|\wh{n_1}(k)|^2}{|k|^2}dk,
\]
For $\la\ge r_\eps^{-4}$ and $k\in B(0,\la^{3/4})$: $|B_\La(k/\la)-B_\La(0)|\le \tfrac{|k|}{\la}(z(\la^{-1/4})+K\La^{-1})$. As $f_1:t\in\mathbf{R}^+\to \tfrac{t}{1+t}$ and $f_2=f_1^2$ have bounded derivatives (by $1$ and $2$ respectively), for $k$ with $B_\La(p)\neq B_\La(0)$,
\[
\begin{array}{ll}
|\WW(k)-\WW(0)|\le \alpha|B_\La(k)- B_\La(0)|,&|\WW(k)^2-\WW(0)^2|\le 2\alpha|B_\La(k)-B_\La(0)|,
\end{array}
\]so
\[
\begin{array}{rl}
\dint_{|k|\le\la^{3/4}}|f_i(\alpha B_\La(k))-f_i(\alpha B_\La(0))|\frac{|\wh{n_\la}(k)|^2dk}{|k|^2}&\le 2\alpha\dfrac{z(\la^{-1/4})+K\La^{-1}}{\la}\dint\dfrac{|\wh{n_1}(k)|^2dk}{|k|}\\
             &\apprle \alpha \dfrac{z(\la^{-1/4})+\La^{-1}}{\la}\ncc{n_1}\nlp{4}{\psi_1}^2.
\end{array}
\]
As $f_1(t),f_2(t)\le t^2$ then 
\[\int_{|k|>\la^{3/4}} \WW(k)^i\frac{|\wh{n_1}(k)|^2}{|k|^2}dk\apprle  \la^{-3/2}L^i\dint |\wh{n_1(k)}|^2dk\apprle \la^{-3/2}L^i\nso{1}{\psi_1}^2=O(L^i\la^{-3/2})
\]
and
\[
\dint_k\WW(k)^i\dfrac{|\wh{n_\la}(k)|^2}{|k|^2}dk=\WW(0)^i\frac{D(n_1,n_1)}{\la}+\underset{\la\to\infty}{o}(L^i\la^{-1}).
\]

There holds $\int_k \alpha^2(1-\WW(k))^2\tfrac{|\wh{\rho}_{1,0}(k)|^2}{|k|^2}dk\apprle \alpha^2\ncc{\rho_{1,0}}^2$. Then estimates of $\ncc{\rho_{1,0}}$ and $\ncc{\sum}$ are proved in Appendix \ref{estimfp}.\hfill{\footnotesize$\Box$}

\noindent\textbf{3. Estimation of $\tr(\D N')$}

We emphasize that $\lpsi$ \emph{has no lower part} as a spinor.

\noindent There holds 
\[
\begin{array}{rl}
\psh{\D \pi\lpsi}{\pi\lpsi}&=-\psh{|\D|\PP\lpsi}{\PP\lpsi}+\psh{\D\g\lpsi}{\g\lpsi}+2\mathfrak{R}\psh{\D\PP\lpsi}{\g\lpsi}\\
                  &= -\psh{|\D|\lpsi}{\lpsi}+O\Big\{\nlp{2}{\g\lpsi}\big(\big\lVert\,|\D| \g\lpsi\big\rVert_{L^2}+\big\lVert\,|\D|\PP \lpsi\big\rVert_{L^2}\big)\Big\}\\
                  &=-\psh{|\D|\lpsi}{\lpsi}+o((L\alpha)^2).
\end{array}
\]
Then we have: 
\[
\begin{array}{rl}
\psh{\D\lpsi}{\pi\lpsi}&=\psh{\D\lpsi}{\g\lpsi}-\psh{|\D|\PPP\lpsi}{\lpsi}\\
                                   &=\psh{|\D| \lpsi}{\PPP \g \PPP \lpsi }+\psh{|\D| \lpsi}{\PPP \g \PP \lpsi}-\psh{|\D|\PPP\lpsi}{\lpsi}\\
                                   &=-\psh{|\D|\PPP\lpsi}{\lpsi}+O(\nlp{2}{\D \lpsi} \ns{2}{\g}^2+\nlp{2}{\D \lpsi}\nb{\g} \nlp{2}{\PP \lpsi})\\
                                   &=-\psh{|\D|\PPP\lpsi}{\lpsi}+o((L\alpha)^2).
\end{array}
\]

Hence $\psh{\D \phi_\la}{\phi_\la}=\frac{\psh{\D \lpsi}{\lpsi}}{1-\nlp{2}{\pi\lpsi}^2}+\psh{|\D| \PP\lpsi}{\lpsi}+o((L\alpha)^2)$.

\begin{notation}
We write $\psh{g_\star\psi}{\psi}$ for $\psh{g_\star(-i\nabla)\psi}{\psi}$ for $\star\in\{0,1\}$.
\end{notation}
As $g_0'(0)=0$ and $\lVert g''_0\rVert_{\infty}\apprle \alpha$ and the $(g'_1)_{\alpha,\La}$'s are uniformly continuous in a neighbourhood of $0$ (\emph{cf} Proposition \ref{g:estimates} in Appendix A)
\[
\begin{array}{r @{$=$} l}
\frac{\psh{\D \lpsi}{\lpsi}}{1-\nlp{2}{\pi\lpsi}^2}&\psh{g_0\lpsi}{\lpsi}(1+\psh{\PP\lpsi}{\lpsi})+o((L\alpha)^2)\\
        & g_0(0)+\frac{g_0(0)}{4}\psh{\frac{g_1^2}{g_0^2}\lpsi}{\lpsi}+o((L\alpha)^2)\\
        & g_0(0)+\frac{g'_1(0)}{4g_0(0)\la^2}\psh{|\nabla|^2\psi_1}{\psi_1}+o((L\alpha)^2).
\end{array}
\]
Furthermore 
\[
\psh{|\D|\PP\lpsi}{\lpsi}=\tfrac{1}{2}\psh{(|\D|-g_0)\lpsi}{\lpsi}=\tfrac{1}{4g_0(0)}\psh{g_1^2\lpsi}{\lpsi}+o(\la^{-2}).
\]
Finally
\begin{equation}
\begin{array}{rll}
\tr(\D N')&=&\psh{\D \phi_\la}{\phi_\la}\\
             &=&g_0(0)+\frac{g_1'(0)^2}{2\la^2g_0(0)}\psh{|\nabla|^2\psi_1}{\psi_1}+o((L\alpha)^2).
\end{array}
\end{equation}

\noindent\textbf{4. Estimation of $\tr(D \g)$}

\begin{notation}
Let us write $B=R'_{\g}-\ph'_{\g}=R(\g+N)-(\rho_\g+n)*|\cdot|^{-1}$.
\end{notation}

\begin{remark}
Let us recall Lemma $1.$\cite{ptf}: if $P,\Pi$ are two projectors such that: 

\noindent$P-\Pi\in\mathfrak{S}_2$ then 

$Q\in\mathfrak{S}_1^P\iff Q\in\mathfrak{S}_1^{\Pi}$ and then $\ttr_P(Q)=\ttr_\Pi(Q)$.
\end{remark}

\noindent We apply this Lemma for $P=\PP$ and $\Pi:=\chi_{(-\infty,0)}(\D+\alpha B)$: formally
\begin{subequations}
\begin{equation}
\tr((\D+\alpha B)\g)=\ttr(|\D|\g^2)+\alpha\tr(B\g)
\end{equation}
\begin{equation}
\tr((\D+\alpha B)\g)=-\ttr(|\D+\alpha B|\g^2)=- \ttr(|\D|\g^2)+o(\ttr(|\D|\g^2)).
\end{equation}
\end{subequations}
So we would like to show that 
\begin{equation}
\begin{array}{rl}
\ttr(|\D|\g^2)&=-\frac{\alpha}{2}\tr(B\g)+o((L\alpha)^2),\\
  &=-\frac{\alpha}{2}\big(D(\rho_\g+n,\rho_\g)-\ttr(R'_\g \g)\big)+o((L\alpha)^2),\\
  &=-\frac{\alpha}{2}D(\rho_\g+n,\rho_\g)+o((L\alpha)^2).
\end{array}
\end{equation}

We have to prove that $B\g$ in $\mathfrak{S}_1^{\PP}$ and $\ttr\big\{ (|\D+\alpha B|-|\D|)\g^2 \big\}=O(\alpha(L\alpha)^2)$.

Supposing those facts are true we get $\ttr(|\D|\g^2)=-\tfrac{\alpha}{2}\tr(B\g)+O(\alpha \tau^2)$. We use \eqref{a3}: 
\[
\ns{1}{R'_{\g} \g}\le \ns{2}{R(\g)|\D|^{-1/2}}\ns{2}{|\D|^{1/2}\g}+\ns{2}{R(N)}\ns{2}{\g}\apprle (\tau+\la^{-1})\tau.
\]

First let us prove that $\ttr(|\D+\alpha B|\g^2)=\ttr(|\D| \g^2)+O(\alpha (L\alpha)^2).$

Thanks to Lemma \ref{squareroot}, there holds:
\begin{equation*}
\left\{
\begin{array}{l}
|\D+\alpha B|\ge |\D|\big(1-\alpha K(\nqkin{\g}+\ncc{\rho_\g}+\nlp{2}{|\nabla|^{1/2} \lpsi}) \big),\\
|\D+\alpha B|\le |\D|\big(1+\alpha K(\nqkin{\g}+\ncc{\rho_\g}+\nlp{2}{|\nabla|^{1/2} \lpsi})\big).
\end{array}
\right.
\end{equation*}
Then we multiply by $\g^*=\g$ on the left and by $\g$ on the right: this does not change the inequalities. To conclude it suffices to take the trace.

\noindent Let us prove $\tr(\ph'_{\g} \g)=D(\rho_{\g}+n_\la,\rho_{\g})$. In fact if $Q\in \mathfrak{S}_1^{\PP}$ and if $\int \ttr(\wh{Q}(p,p))dp$ exists then this last integral is equal to $\tr(Q)$, because $\PP=f(i\nabla)$, in Fourier space we have $\ttr_{\CC}(\wh{\PP}(p) \wh{Q}(p,p) \wh{\PPP}(p))=\ttr_{\CC}(\wh{\PPP}(p) \wh{Q}(p,p) \wh{\PP}(p))=0.$ Here, the trace $\tr(\ph'_{\g} \g)$ is formally equal to
\[
\begin{array}{l}
(2\pi)^{-3/2}\!\diint\limits_{|p|,|q|<\La} \wh{\ph'_{\g}}(p-q)(\ttr(\wh{\g}(p,q)))^*dpdq\\
\ \ \ \ \ \ \ \ \ \ \ \ \ \ \ \ =(2\pi)^{-3/2}\!\diint\limits_{|u+\tfrac{k}{2}|,|u-\tfrac{k}{2}|<\La} \wh{\ph'_{\g}}(k)(\ttr(\wh{\g}(u+k/2,u-k/2)))^*dudk\\
\ \ \ \ \ \ \ \ \ \ \ \ \ \ \ \ =\dint_k  \wh{\ph'_{\g}}(k) \wh{\rho_\g}(k)^*dk=4\pi\dint_k \frac{\wh{\rho'_\g}(k)\wh{\rho_\g}(k)^*}{|k|^2}dk=D(\rho_\g,\rho'_\g).
\end{array}
\]
As shown in the estimation of $I$, there holds
\[
D(\rho_\g,\rho_\g+n_\la)=\tfrac{\WW(0)^2-\WW(0)}{\la}D(n_1,n_1)+o\big(\tfrac{L}{\la}\big),
\]
so we get
\begin{equation}
\ttr(|\D|\g^2)=\alpha\tfrac{\WW(0)^2-\WW(0)}{2\la}D(n_1,n_1)+o\left(\frac{L}{\la}\right).
\end{equation}

\begin{remark}
The calculation above is correct if $\wh{\g}(p,q)\in\mathcal{C}^0(B(0,\La)^2)$:
\[
\diint\limits_{|u\pm\tfrac{k}{2}|<\La}\frac{|\wh{\rho}(k)|}{|k|^2}|\wh{\g}(u+\tfrac{k}{2},u-\tfrac{k}{2})|dudk \apprle \La^{3/2}\ncc{\rho}(\La^{3/2}\nlp{\infty}{\wh{\g}}+\ns{2}{\g}) .
\]
We conclude by continuity of $Q\in\mathfrak{S}_1^{\PP}\mapsto \rho_Q\in\mathcal{C}$ shown in \cite{at}, that of 

\noindent $Q\in\mathfrak{S}_1^{\PP}\mapsto \tr(\ph'_{\g} Q)$ and the density of $\mathcal{C}^0(B(0,\La)^2)$ in $\mathscr{F}(\mathfrak{S}_1^{\PP}(\hl))$. 

Let us prove $\ph'_\g Q\in\mathfrak{S}_1^{\PP}$. We have:
\begin{equation}\label{chu}
(\ph'_{\g} Q)^{--}=\underbrace{(\PP[\ph'_{\g},\PPP]|\D|^{-1/2})}_{\in\mathfrak{S}_2(\hl)}\underbrace{|\D|^{1/2}Q^{+-}}_{\in\mathfrak{S}_2(\hl)}+\underbrace{(\ph'_{\g} |\D|^{-1/2})^{--}}_{\in\mathcal{B}(\hl)} \underbrace{|\D|^{1/2}Q^{--}}_{\in \mathfrak{S}_1(\hl)}\in \mathfrak{S}_1(\hl)
\end{equation}
and so $|\tr(\ph'_{\g} Q)|\le (\La^{1/2}+\sqrt{\llo})\ncc{\rho'_{\g}} \big\lVert Q\big\rVert_{\mathfrak{S}_{1,\PP}}$ with
\begin{equation}\big\lVert Q\big\rVert_{\mathfrak{S}_{1,\PP}}:=\ns{1}{Q^{--}}+\ns{1}{Q^{++}}+\ns{2}{Q^{-+}}+\ns{2}{Q^{+-}}.
\end{equation}

To see $\PP[\ph'_{\g},\PPP]|\D|^{-1/2}$ is Hilbert-Schmidt, it suffices to prove the kernel of its Fourier transform is in $L^2(B(0,\La)^2)$: this is easy with the help of Lemma \ref{m(p,q)}.
\end{remark}

\noindent To conclude this section there remains to deal with $R'_\g\g$, we recall this operator is trace-class (\emph{cf} Lemma \ref{rrq}):
\[
R_\g'\g=\underbrace{(R_\g') |\D|^{-1/2}}_{\in\mathfrak{S}_2(\hl)}\underbrace{|\D|^{1/2}\g}_{\in\mathfrak{S}_2(\hl)}
\]
and
\begin{equation}
\ttr\big\{ R'_\g\g\big\}=O(\ns{2}{R_N}\ns{2}{\g}+\ns{2}{R_\g |\D|^{-1/2}}\ns{2}{|\D|^{1/2}\g})=O\Big(\frac{\alpha\sqrt{L\alpha}}{\la}+(L\alpha)^2\Big).
\end{equation}

\begin{remark}
As $\La\to+\infty$ there holds $\psh{|\D|^2\psi_1}{\psi_1}-D(n_1,n_1)=E_{\text{CP}}+o(1)$. In fact $\psi_1=(\phi_1, 0)^{T}$ where $\phi_1=P_{\hl}\phi_1'/\nlp{2}{P_{\hl} \phi_1'}$ and $\phi'_1$ is the minimizer of Choquard-Pekar energy. $P_{\hl}$ is the projector onto $\hl$. So we have $\phi_1\overset{H^1}{\underset{\La\to+\infty}{\to}}\phi'_1$. Writing $n'=|\phi'_1|^2$ there holds by Kato's inequality~\eqref{kato}
\[
\begin{array}{rl}
\big|\ncc{n_1}-\ncc{n'}\big|&\le \ncc{n_1-n'}\apprle \Big(\psh{|\nabla| \psi_1}{\psi_1}+\psh{|\nabla|\phi'_1}{\phi'_1}\Big)\big|\nlp{2}{\psi_1}^2-\nlp{2}{\phi'_1}^2\big|\\
                                  &\apprle \psh{|\nabla|\phi'_1}{\phi'_1}\Big|\nlp{2}{\psi_1}^2-\nlp{2}{\phi'_1}^2\Big|\underset{\La\to\infty}{\to} 0.
\end{array}
\]
\end{remark}

\subsection{Proof of Proposition \ref{HVZ}}\label{proofHVZ}
In this part we write $E(\cdot)$ for $E_{\text{BDF}}^0(\cdot)$.

Let us prove now the binding inequalities for $0<q<1$. According to Lieb's principle (Proposition 3.\cite{at}) for each $q$ we can take minimizing sequences for $E(q)$ of the form
\begin{equation}
\left\{
\begin{array}{l}
Q_{(k)}=P_{(k)}-\PP+q\ket{\psi_k}\bra{\psi_k},k\in\mathbf{N}\\
\text{with\ } (P_{(k)}-\PP)\in\mathcal{Q}_\La(0)\text{\ and\ } P_k^2=P_k,P_k\psi_k=0.
\end{array}\right.
\end{equation}
We write as before $\g_k=P_k-\PP,n_k=|\psi_k|^2,N_k=\ket{\psi_k}\bra{\psi_k}$. We will forget to emphasize the dependence in $k$.\newline
\noindent Writing $I_\g(N)=\alpha\mathfrak{R}\Big(D(\rho_\g,n)-\ttr(R_N^* \g)\Big)$, $\EE(Q)$ can be written:
\[\begin{array}{l}
\EE(Q)=\EE(\g)+q\psh{\D\psi}{\psi}+qI_\g(N)=(1-q)\EE(\g)+q\EE(\g+N).
\end{array}\]
Taking the $\liminf$, we obtain
\[
E(q)=\liminf\limits_{k\to\infty}((1-q)\EE(\g)+q\EE(\g+N))\ge (1-q)\liminf\limits_{k\to\infty}\EE(\g)+qE(1).
\]
Either $x=\liminf\limits_{k\to\infty}\EE(\g)>0$ and $E(q)>qE(1)$ or $x=0$. What happens in the second case ? Up to the extraction of a subsequence we can assume that $\liminf \EE(\g)$ is a limit. Thanks to \eqref{eqin} it implies $\ttr(|\D|\g^2)+D(\rho_\g,\rho_\g)\underset{k\to\infty}{\to} 0$. As $P_k\psi_k=0$, there holds $\PP \psi_k=\g\psi_k$, in particular 
\[
\lVert\PPP \psi\rVert^2=\lVert\psi\rVert^2-\lVert\PP\psi\rVert^2=1-\lVert\g\psi\rVert^2\to 1
\]
and $\psh{\D \psi}{\psi}=\psh{|\D|\psi^+}{\psi^+}-\psh{|\D|\g\psi}{\g\psi}$ where $\psi^{\eps}=\mathcal{P}^0_{\eps}\psi$. 

\noindent As $\nlp{2}{|\D|^{1/2}\g\psi}^2\le \ttr(|\D|\g^2)\nlp{2}{\psi}^2$ and $\lVert\psi\rVert_2=1$: up to extraction we have
\[
\lim\limits_{k\to\infty}\psh{\D\psi}{\psi}=\lim\limits_{k\to\infty}\psh{|\D|\psi^+}{\psi^+}\ge m(\alpha).
\]
The sequence $(\psh{\D\psi_k}{\psi_k})_k$ is bounded, else by Cauchy-Schwarz and Kato's inequality
\[
\begin{array}{rl}
\EE(\g+N)&\ge \EE(\g)+\psh{\D\psi}{\psi}-\dfrac{1}{2}\big(\ncc{\rho_\g}^2+\nqq{\g}^2+\pi \alpha^2\psh{|\nabla| \psi}{\psi} \big)\\
      &\underset{k\to+\infty}{\longrightarrow}+\infty.
\end{array}
\]
By Cauchy-Schwartz inequality $I_\g(N)\to 0$ and 
\[
\liminf_{k\to\infty}\EE(Q_k)=E(q)\ge \liminf_{k\to\infty}\EE(\g)+q\liminf_{k\to\infty} I_\g(N)+q\liminf_{k\to\infty} \psh{\D\psi}{\psi}\ge qm(\alpha).
\]
It implies $E(q)=qm(\alpha)$, but we can use the method of Section \ref{pllsup}. to prove that $E(q)<qm(\alpha)$ for sufficiently small $\alpha$ and $L$ in regard with $q$: we define $\ov{Q}$ by the formulae
\[\left\{
\begin{array}{l}
\ov{\Pi}:=\ov{\g}+\PP=\chi_{(-\infty,0)}\Big(\D+\alpha\big(\ph_{\ov{\g}}+qn*|\cdot|^{-1}-R(\ov{\g}+qN)\big)\Big),\\
\ov{Q}:=\ov{\g}+\frac{q}{1-\nlp{2}{\ov{\Pi}\lpsi}^2}\ket{(1-\ov{\Pi})\lpsi}\bra{(1-\ov{\Pi})\lpsi}.
\end{array}\right.
\]
If we assume that $E(q)=qm(\alpha)$ once $E(1)<m(\alpha)$ has been proven, we also obtain $E(q)>qE(1)$. We thus get $E(q)+E(1-q)>qE(1)+(1-q)E(1)=E(1)$.

There remains the case $q>1$. However it has been proved in \cite{at} that for each integer $M$, $E_{\text{BDF}}(\cdot)$ is concave on $[M,M+1]$. Besides thanks to \eqref{eqin} there holds

\[E(q)\ge q(1-\alpha\tfrac{\pi}{4})m(\alpha).\]
\noindent So it suffices that $2(1-\alpha\tfrac{\pi}{4})m(\alpha)>E(1)$ to get $E(q)>E(1)$ for $q>1$. For $\alpha<\tfrac{2}{\pi}$ it is true and as $E(q)>0$ for $q\neq 0$ the binding inequalities for $q>1$ are proved.

\subsection{Proof of Theorems \ref{tlinf} and \ref{nrlim}}\label{end?}

\subsubsection{Notations}\label{mu}

Let $Q=\g'=\g+N$ be a minimizer written with the notation of \eqref{eqg}. As before we write $n:=\rho_N$.

%[
We have $N=\chi_{(0,\mu]}(D_Q)$ with $D_Q:=\D+ \alpha (R'_\g-\ph'_\g)$.  We have to show that $N=\ket{\psi}\bra{\psi}$, then we can choose $\mu$ such that $D_Q\psi=|D_Q|\psi=\mu \psi$ with $\mu\le m(\alpha)$.

We split $\psi$ in two: $\psi=\begin{pmatrix} \ph\\ \chi\end{pmatrix}$. The wave function $\ph\in L^2(\RR,\mathbf{C}^2)$ is the upper spinor and $\chi\in L^2(\RR,\mathbf{C}^2)$ the lower spinor.
%)

We write $C_0^2:=\tfrac{2g'_1(0)^2}{(\alpha \WW(0))^2m(\alpha)}$ and $c:=\tfrac{(g'_1(0))^2}{\alpha \WW(0)m(\alpha)}$.

As $(R(N)-\psi_{|\psi|^2})\psi=0$, there holds
\begin{equation}\label{equ}
(\D+\alpha (\ph_\g-R_\g))\psi=\mu\psi=|\D+\alpha (\ph_\g-R_\g)|\psi.
\end{equation}

We write $v_\g^{\star}:=\ph^{\star}_\g,b_\g^{\star}:=v_\g^{\star}-R_\g^{\star}$, where $\star$ is a prime symbol or no prime. Moreover we write $d:=\D$. We recall:
\begin{equation}\label{remarque}
\psh{v_\g\psi}{\psi}=D(\rho_\g,n)\text{\ and\ }|\psh{R_\g\psi}{\psi}|\le \nqq{\g}\ncc{n},
\end{equation}

We recall the notation $\psh{g_\star\psi}{\psi}:=\psh{g_\star(-i\nabla) \psi}{\psi}$ with $\star\in\{0,1\}$.

\subsubsection{Strategy of the proof}

The proof of Theorem \ref{tlinf} relies on bootsrap arguments enabling us to get appropriate estimates of $\nlp{2}{\,|\nabla|^s \psi}$ for $s=\tfrac{1}{2},1,\tfrac{3}{2}$.
The starting point is \emph{a priori} estimates of $\nlp{2}{\,|\nabla|^{1/2} \psi},\nlp{2}{\nabla \psi},\ttr(|\D| \g^2)$.
It is possible to use an adaptation of the fixed point method of \cite{ptf} to get estimates of
\[
\diint \ed{p-q}^{2s}\ed{p+q}|\wh{g}(p,q)|^2dpdq\text{\ and\ }\dint \frac{\ed{k}^{2s}|\wh{\rho}_\g(k)|^2}{|k|^2}dk
\]
in terms of the Sobolev norms $\nso{s+1/2}{\psi}$ at least for $s=0,\tfrac{1}{2},1$. Then the second part of Eq.~\eqref{eqg} enables to get estimates of $\nlp{2}{\,|\nabla|^{s+1} \psi}$ in terms of $\nlp{2}{\,|\nabla|^{s+1/2} \psi}$ and the (squared) norms above. It is possible to keep going as explained in the thesis of the author \cite{these}, provided $\alpha,L$ are small enough.

More precisely the steps are the following.

\noindent 1. We first prove \emph{a priori} estimates and get $\ncc{\rho_Q},\nqkin{Q}$ are $O(1)$ and then show that $\nqkin{\g}=o(1)$. As a consequence Lemma \ref{happens} holds and we get $\ncc{\rho_\g}, \nqo{\g},\psh{|\D|\psi}{\psi}$ are O(1). This enables us to show that we can apply the fixed point method (Lemma \ref{fpqu}, $f=1$) and that the minimizer $\g+N$ and its density $\rho_\g+n$ form a fixed point (at least in the space associated to $\nqo{\cdot}$ and $\ncc{\cdot}$).

\noindent 2. We then prove
\begin{equation}\label{est1}
\nso{3/2}{\psi},\nc{n}=O(1).
\end{equation}
\noindent Thus we can apply the fixed-point method (Lemma \ref{fpq}) with $n=|\psi|^2$ and $N=\ket{\psi}\bra{\psi}$ and so to construct $(\g+N;\rho_\g+\rho_N)$ as a fixed point in (a ball of) $\mathcal{X}$.

\noindent 3. Using the estimates that we deduce from the fixed-point method and Eq.~\eqref{eqg} we then prove that
\begin{equation*}\label{H1dl}
\psh{|\nabla|^2\psi}{\psi}=O((\alpha \WW(0))^2).
\end{equation*}

\noindent 4. Following \cite{at}, we apply a scaling transform to the minimizer with the scaling factor $c=O(\alpha \WW(0))$ defined in Subsection \ref{mu}: 

\noindent we get $\un{\psi}(x):=c^{3/2}\psi(cx)\in H^1(\CC)$. The previous results will give
\begin{equation*}\label{est3}
\nso{3/2}{\un{\psi}}=O(1),\ \lVert\un{\chi}\rVert_{H^1}=O(L\alpha),
\end{equation*}
where $\un{\chi}\in H^1(\RR,\mathbf{C}^2)$ is the lower spinor of $\un{\psi}$.

\noindent 5. At last we compute the energy and show the asymptotic expansion.

\subsubsection{\emph{A priori} estimates}
The first step is the following result.
\begin{lemma}\label{1st_step}
For $Q=\g+N$ a minimizer of $E_{\text{BDF}}^0(1)$, then $N$ has rank 1 and there holds the following $\emph{a priori}$ estimates:
\[
\ttr(|\D|\g^2)+\alpha D(\rho_\g,\rho_\g)+\psh{|\D|\psi}{\psi}\apprle 1.
\]
The decomposition $\g+N$ is the same as in \eqref{eqg}, Section \ref{main} with $N=\ket{\psi}\bra{\psi}$.
\end{lemma}
Assuming this result is true we can go further: we know that 

\noindent $F(Q,\rho_Q)=(Q,\rho_Q)$ where $F$ is the function defined in \eqref{cauchy1} and \eqref{cauchy2}. Using the estimates of Appendix \ref{estimfp} we get that:
\[
 \ncc{\rho_\g}\apprle L\ncc{n}+\sqrt{L\alpha}\nqo{Q}+\ssum_{j=2}^{+\infty}(\alpha K(\ncc{\rho_Q}+\nqo{Q}))^j\apprle L=O(1). 
\]
We then apply Lemma \ref{fpqu} (with $f=1$): we get that $(Q,\rho_Q)$ is in fact the \emph{unique} fixed point of $F$ in a ball of $\mathcal{X}_0$.

\noindent\textbf{Proof of Lemma\ }\ref{1st_step}:
As $Q$ is a minimizer and that $E_{\text{BDF}}^0(1)\le m(\alpha)$ then there holds:
\begin{equation}
m(\alpha)\ge \EE(Q)\ge \big(1-\alpha \frac{\pi}{4}\big)\ttr(|\D|Q^2)+\frac{\alpha}{2}D(\rho_Q,\rho_Q),
\end{equation}
and $\nqkin{Q},\sqrt{\alpha}\ncc{\rho_Q}=O(1)$. As $\g=\chi_{(-\infty,0)}(\D_Q)-\PP$, using estimates of Lemmas \ref{lin} and \ref{nonlin} we get:
\[
\ns{2}{\g}\apprle \alpha(\ncc{\rho_Q}+\nqq{Q})\apprle \sqrt{\alpha}.
\]
Thus $|\tr(\g)|\le \ns{2}{\g}^2\apprle \alpha <1$, as a consequence $\tr(\g)=0$ and $N$ has rank 1.

Thanks to \eqref{remarque} and Kato's inequality there holds
\begin{equation}
\psh{\D \psi}{\psi}=\psh{D_Q \psi}{\psi}-\alpha \psh{b_\g\psi}{\psi}=\psh{|D_Q|\psi}{\psi}+O(\alpha \nso{1/2}{\psi}(\nqq{\g}+\ncc{\rho_\g})).
\end{equation}
We apply Lemma \ref{squareroot} on $|D_Q|$:
\begin{equation}
\psh{|D_Q|\psi}{\psi}\ge \big(1- K(\alpha\nqq{Q}+\alpha^{1/2}\times\alpha^{1/2}\ncc{\rho_Q})\big)\psh{|\D|\psi}{\psi},
\end{equation}
and: 
\begin{equation}
\psh{\D \psi}{\psi}\ge (1-K\sqrt{\alpha})\psh{|\D|\psi}{\psi}+O\big(\alpha(\ncc{\rho_Q}+\nqkin{Q})\nso{1/2}{\psi}\big).
\end{equation}

\noindent By Cauchy-Schwartz inequality and Kato's inequality:
\[
\begin{array}{rl}
\EE(Q)&=\EE(\g)+\psh{\D\psi}{\psi}+\alpha\mathfrak{R}\big(D(\rho_\g,n)-\ttr( R_\g N)\big)\\
       &\ge (1-\alpha\tfrac{\pi}{4})\ttr(|\D|\g^2)+\frac{\alpha}{2}D(\rho_\g,\rho_\g)+(1-C_2\sqrt{\alpha})\psh{|\D|\psi}{\psi}\\
       &\ \ \ \ \ \ \ \ \ \ \ \ -\alpha \nso{1/2}{\psi}(\nqkin{\g}+\ncc{\rho_\g}).
\end{array}
\]
As $\EE(Q)\le m(\alpha)$ we have 
\begin{equation}\label{gros1}
\ttr(|\D|\g^2)+\alpha D(\rho_\g,\rho_\g)+\psh{|\D|\psi}{\psi}=O(1).
\end{equation}
\hfill{\footnotesize$\Box$}

\subsubsection{Estimates around the fixed point method}
 Let us prove that we can construct $(Q,\rho_Q)$ as a fixed point in $\mathcal{X}$. We have to show $\nc{n},\nq{N}=O(1)$ and as $\nq{N}\apprle \nso{3/2}{\psi}^2$ it suffices to prove  \eqref{est1}.

By Sobolev  inequality \eqref{sobin}: 
\[
\nlp{2}{n}=\nlp{4}{\psi}^2\apprle \big\lvert\big\vert\,|\nabla|^{3/4}\psi\big\rVert_{L^2}^2\apprle\nlp{2}{\nabla \psi}^{3/2}\nlp{2}{\psi}^{1/2}=O(\sqrt{\alpha}).
\]
Moreover there holds $D(n,n)\le \tfrac{\pi}{2}\psh{|\nabla|\psi}{\psi}\apprle 1$ and $\nc{n}=O(1)$.

At this point we have: $\ncu{n},\nqu{N}\apprle 1$: we can apply Lemma \ref{fpqu} with

\noindent $f(p-q)=\ed{p-q}$ and construct $(Q,\rho_Q)$ as a fixed point in $\mathcal{X}_1$. As shown in Appendix \ref{estimfp}, there holds $\ncu{\g}+\ncu{\rho_\g}\apprle 1$.

Let us now prove that $\nso{3/2}{\psi}\apprle 1.$ By \eqref{equ} we have $|d|^2\psi=\mu d\psi-\alpha db_\g\psi$, therefore:
\[
\psh{|d|^3\psi}{\psi}=\mu \psh{|d|d\psi}{\psi}+\alpha \psh{|d|^{1/2}(R_\g-v_\g)|d|^{-3/2}|d|^{3/2}\psi}{d|d|^{1/2}\psi}.
\]

Then thanks to \eqref{a2} and Lemma \ref{chiant} below, writing
\[
|d|^{1/2}b_\g|d|^{-3/2}=[\,|d|^{1/2},b_\g]|d|^{-3/2} +b_\g|d|^{-1}
\]
we get $\big\lVert |d|^{1/2}b_\g|d|^{-3/2}\big\rVert_{\mathcal{B}}\apprle (\nqq{\g}+\ncc{\rho_{\g}})+\nqu{\g}$. 

We obtain at last $\nso{3/2}{\psi}\apprle 1$. In particular we can apply Lemma \ref{fpq} and construct $(Q,\rho_Q)$ as a fixed point in $\mathcal{X}$ and get $\nq{\g},\nc{\rho_\g}\apprle 1$.

\begin{lemma}\label{chiant}
Let $(\g'_0,\rho'_0)$ be in $\mathcal{Q}\times \mathcal{C}$ and $b_0:=\rho'_0*\tfrac{1}{|\cdot|}-\g'_0,v'_0:=\rho'_0*\tfrac{1}{|\cdot|}.$ Then there holds:
$\bigg|\bigg|\,|\D|^{-\tfrac{3}{2}}\Big[b_0,|\D|^{\tfrac{1}{2}}\Big]\bigg|\bigg|_{\mathcal{B}}+\bigg\lVert |\D|^{-1}\Big[b_0,|\D|^{\tfrac{1}{4}}\Big]\bigg\rVert_{\mathcal{B}} \apprle (\nq{\g'_0}+\ncc{\rho'_0}).$
\end{lemma}
\noindent\textbf{Proof}:\ The estimation for the term $R(\g'_0)$ comes from \eqref{a2} in Lemma \ref{rrq}: indeed we have
\[
|\ed{p}^{s}-\ed{q}^{s}|\le K\frac{|p-q|}{\ed{p}^{1-s}+\ed{q}^{1-s}}\text{for\ } s=\dfrac{1}{2^k},\,k\in\mathbf{N}^*.
\] 
We write $f\in \hl$ and $\Phi=|\D|^{-\tfrac{3}{2}}\Big[v'_0,|\D|^{\tfrac{1}{2}}\Big]$, the following holds:
\[
\dint_p |\wh{\Phi f}(p)|^2dp\le K\diint \frac{dpdq}{\ed{p}^3}\frac{|\ed{p}-\ed{q}|^2}{|p-q|^4}\frac{|\wh{\rho'_{0}}(p-q)|^2}{\ed{p}+\ed{q}}\dint |\wh{f}(q)|^2dq.
\]
To deal with last term we use the same method.\hfill{\footnotesize$\Box$}

\noindent Let us prove $\psh{|\nabla|^2\psi}{\psi}=O((\alpha \WW(0))^2).$

We write $x=x(N)=\nlp{2}{g_1(-i\nabla) \psi}$. By Lemma \ref{noir} we have:
\begin{subequations}
\begin{equation}
\ncc{\rho_\g}\apprle  Lx^{1/2}+\alpha x+L\alpha,
\end{equation}
\begin{equation}
\nqq{\g}\apprle \sqrt{L\alpha} x^{1/2}+\alpha x+L\alpha.
\end{equation}
\end{subequations}

Taking $\nlp{2}{\cdot}$-norm of $d \psi=\mu\psi-\alpha b_\g\psi$, we have (\emph{cf} Proposition \ref{g:estimates} for $\lVert g''_0\rVert_{\infty}$):
\[
\begin{array}{l}
\psh{d^2 \psi}{\psi}=x^2+m(\alpha)^2+O(\lVert g''_0\rVert_{\infty} x^2)=x^2+m(\alpha)^2+O(\alpha x^2)\\
\alpha |\psh{b_\g \psi}{\psi}|+\alpha^2\nlp{2}{b_\g\psi}^2\le K_1L\alpha^{2}x^{1/2}+K_2L\alpha x+K_3\alpha^2 x^{3/2}+K_4(L\alpha^3)x^2+K_6\alpha^4x^{3}\\
\mu^2\nlp{2}{\psi}^2\le m(\alpha)^2.
\end{array}
\]
For the first equality we have used Taylor's Formula (order 2) and the fact that $g'_0(0)=0$. As $x=O(1)$ we have $\alpha^4x^3=O(\alpha^{4}x^2)$ and
\begin{equation}
x^2\le k_1L\alpha^{2}x^{1/2}+k_2 (L\alpha)x+k_3 \alpha^2 x^{3/2}.
\end{equation}
Finally we obtain
\begin{equation}
x^{1/2}\le k_1^{1/3}(L\alpha^2)^{1/3}+k_2^{1/2}(L\alpha)^{1/2}+k_3\alpha^2\apprle (L\alpha)^{1/2},
\end{equation}
and there holds $x^2\le K(L\alpha)^2=O(c^{-2})$.

By Lemma \ref{noir} the following estimates hold for the minimizer:
\[
\begin{array}{l|l}
\nq{\g}\apprle \alpha, & \nc{\rho_\g}\apprle (L+w(N))\sqrt{L\alpha},\\
\nhi{\g}\apprle  L\alpha, & \ncc{\rho_\g}\apprle L\sqrt{L\alpha}\\
\end{array}
\]
where we recall:
\[
w(N):=\bigg\{\diint |p-q|^2|p+q| |\wh{N}(p,q)|^2dpdq\bigg\}^{1/2}\apprle \big\lVert\,|\nabla|^{3/2} \psi\big\rVert_{L^2}.
\]

\subsubsection{Scaling}

\noindent We have considered so far the problem associated with $E_{c=1,\alpha,\La}$(BDF energy where the parameters are: speed of light $1$, fine structure constant $\alpha$ and cut-off $\La$). We link it to the BDF energy in another choice of parameters: speed of light $c$, fine structure constant $\alpha c$ and cut-off $c\La$, with $c>0$ defined in Subsection \ref{mu}.

As in \cite{at} we write
\[
U_c^*: \begin{array}{rll}
\hl& \to& \mathfrak{H}_{c\La} \\
\phi&\mapsto & c^{3/2}\phi(c(\cdot)),
\end{array}
\] 
and so $U_c\phi(x)=c^{-3/2}\phi(x/c)$. There holds a scaling correspondence between $(1,\alpha,\La)$ and $(c,c\alpha,c\La)$ :
\[
E_{c,c\alpha,c\La}(U_c^*QU_c)=c^2E_{1,\alpha,\La}(Q).
\]
\noindent To distinguish the corresponding objects of $(c,c\alpha,c\La)$ we underline them:
\[
\begin{array}{l | l}
\un{\psi}(x)=U_c^* \psi(x)= c^{3/2}\psi(cx), &\un{\D}=c^2 \, U_c^*\D U_c=\un{m}c^2\beta+cT,\\
\un{\g}(x,y)=U_c^*\g U_c(x,y)=c^3\g(cx,cy),&\un{m}=g_0(-i\nabla/c),\\
\rho_{\un{\g}}(x)=c^3\rho_\g(cx),\un{v}=\rho_{\g}*|\cdot|^{-1} ,&T_{\boldsymbol{\alpha}}=cg_1(-i\nabla/c)\boldsymbol{\alpha}\cdot \frac{-i\nabla}{|\nabla|},\\
\un{R}(x,y)=\un{\g}(x,y)|x-y|^{-1},&T_{\sigma}=cg_1(-i\nabla/c)\sigma\cdot \frac{-i\nabla}{|\nabla|}.
\end{array}
\]
\noindent There holds $|\nabla|\le |\tigma|\le C_1 |\nabla|$ and 
\[
\left\{
\begin{array}{l}
\nqq{\un{\g}}=\sqrt{c}\nqq{\g}\\
\ncc{\rho_{\un{\g}}}=\sqrt{c}\ncc{\rho_\g}
\end{array}\right.,\text{so}
\left\{
\begin{array}{l}
\nb{\un{R}|D^0|^{-1/2}}\apprle \nqq{\g}= \sqrt{c}\nqq{\g}\\
\nb{\un{v}|D^0|^{-1/2}}\apprle \ncc{\rho_{\un{\g}}}=\sqrt{c}\ncc{\rho_\g}
\end{array}
\right.
\]

\noindent We have shown $\psh{g_1^2\psi}{\psi}=O((L\alpha)^2)$, so for $c:=\tfrac{g'_1(0)^2}{\alpha \WW(0)}$, $\un{\psi}$ has uniformly bounded $H^1$ norm with respect to the parameters in the regime \eqref{regi}.

\begin{remark}
Here the constant of scaling $c$ corresponds to $\la$ of the test function.
\end{remark}

First we we prove the following middle results.
\begin{lemma}
Let $Y=Y(\psi):=\nlp{2}{g_1^{3/2} \psi}$ where $\psi$ is defined as above. Then we have
\[
\nlp{2}{\chi}\apprle c^{-1}\text{\ and\ } \nlp{2}{\nabla \chi}\apprle \alpha Y+c^{-1}.
\]
Moreover $\mu=m(\alpha)+O(c^{-2})$ and $E_{\text{BDF}}^0(1)=\EE(\g')=m(\alpha)+O(c^{-2})$.
\end{lemma}
\noindent\textbf{Proof}: Thanks to \eqref{equ} we have
\begin{equation}
\un{m}c^2\beta\unp+c\talpha \unp +\alpha c (\un{v}-\un{R})\unp=\mu c^2\unp.
\end{equation}

Considering the upper part  $\ph$ and the lower part $\chi$ of $\psi$:
\begin{subequations}
\begin{equation}\label{up}
\un{m}c^2\un{\ph}+c\tigma\un{\chi}+\alpha c \un{v}\un{\ph}-\alpha c(\un{R}\unp)_1 =\mu c^2\un{\ph}
\end{equation}
\begin{equation}\label{low}
-\un{m}c^2\chi+c\tigma\ph+\alpha c \un{v}\chi-\alpha c(\un{R}\unp)_2= \mu c^2\un{\chi}
\end{equation}
\end{subequations}

From \eqref{low} we obtain
\[
\un{\chi}=\frac{\tigma}{\un{m}c+\mu c}\un{\ph}+\frac{\alpha}{\un{m}c+\mu c}((\un{R}\unp)_2-\un{v}\un{\chi}).
\]
We take the $L^2$-norm:
\[
\lVert\un{\chi}\rVert_{L^2}\apprle \frac{\lVert\unp\rVert_{H^1}}{c}+\frac{\alpha}{\sqrt{c}}(\ncc{\rho_\g}+\nqq{\g})\apprle \frac{1}{c}+\frac{\alpha L\sqrt{L\alpha}}{\sqrt{c}}+\frac{\alpha L\alpha}{\sqrt{c}}\apprle\frac{1}{c}.
\]
In particular we have $\nlp{2}{\chi}=\nlp{2}{\un{\chi}}=O(c^{-1})$.

We write $S_{\mathbf{x}}=\boldsymbol{g}_1(-i\nabla)\cdot \mathbf{x}$ with $\mathbf{x}$ either $\sigma$ or $\boldsymbol{\alpha}$. As $\talpha$ exchanges upper and lower spinors, by Cauchy-Schwarz inequality the following holds:
\[
\begin{array}{rl}
\psh{\D \psi}{\psi}&=\psh{g_0 \ph}{\ph}-\psh{g_0 \chi}{\chi}+2\mathfrak{R}\psh{S_\sigma\ph}{\chi}\\
                              &= m(\alpha) \nlp{2}{\ph}^2+O(c^{-2})\\
                              &=m(\alpha) +O(c^{-2}).
\end{array}
\]

It enables us to estimate 
\begin{equation}\label{muE}
\mu=m(\alpha)+O(c^{-2})\text{\ and\ }E_{\text{BDF}}^0(1)=\EE(\g')=m(\alpha)+O(c^{-2}).\end{equation}

From Eq.~\eqref{up} we get
\[
\tigma\un{\chi}=\frac{(\mu c^2-\un{m}c^2)\un{\ph}}{c}+\alpha[(\un{R}\psi)_1-\un{V}\un{\ph}].
\]
As $\mu=m(\alpha)+O(c^{-2})$, the $L^2$-norm of $\tigma\un{\chi}$ has the following upper bound:
\[
\lVert\tigma\un{\chi}\rVert_{L^2}\apprle \alpha+c^{-1}+\alpha\sqrt{c}(L\alpha+L\sqrt{L\alpha})\apprle \alpha,
\]

writing $Y^2=Y(\psi)^2:=\psh{g_1^3\psi}{\psi}$, we get the middle estimates
\begin{subequations}
\begin{equation}\label{gros3}
\lVert\un{\chi}\rVert_{H^1}\apprle \alpha
\end{equation}
\begin{equation}
\nso{1}{\un{\chi}}\apprle (\alpha Y+c^{-1}).
\end{equation}
\end{subequations}

Indeed writing $\mu=m(\alpha)+\delta m$, $c^2\times\frac{\delta m}{c}\un{\ph}$ has $L^2$-norm lesser than $Kc^{-1}$. Then:
\[
\Big|g_0(p/c)-g_0(0)\Big|=\left\{
\begin{array}{l l}
 \Big|\dint_0^1 g'_0(tp/c)dt\frac{|p|}{c}\Big| &\le K\alpha\frac{|p|}{c},\\
 \Big| \dint_0^1 g''_0(tp/c)(1-t)dt\frac{|p|^2}{c^2}\Big| &\le K\alpha \frac{|p|^2}{c^2},\end{array}\right.
\]

and $|g_0(p/c)-g_0(0)|\apprle \alpha|p|^3/c^3$.  In particular

\begin{equation}\label{guchi}
\psh{g_1\chi}{\chi}\le\sqrt{\psh{\chi}{\chi}\psh{g_1^2\chi}{\chi}}=O(c^{-1}\times (\alpha Y+c^{-1})c^{-1})=O(\alpha Yc^{-2}+c^{-3})
\end{equation}
and there also holds the middle estimate: $\nlp{2}{\nabla \chi}\apprle \alpha c^{-1}$.\hfill{\footnotesize$\Box$}

\noindent Let us prove that $\nso{3/2}{U_c^*\psi}=O(1).$ The method is the following: we take the scalar product of $|\nabla| \psi$ with each part of the equation $|\D|^2\psi=\D(\mu-\alpha b_\g)\psi$. Then we cancel the leading terms in order to get an inequality involving $Y^2=\psh{g_1^3 \psi}{\psi}$ of the form:
\[
Y^2\le O(c^{-3}+Yc^{-3/2}+Y^{3/2} c^{-3/4}).
\]

As a consequence we get $Y^2=O(c^{-3})$.

\noindent Let us first deal with $\psh{|\D|^2 \psi}{|\nabla|\psi}$.

Thanks to estimate \eqref{guchi} there holds
\[
\Big|\mu\psh{g_1\boldsymbol{\alpha}\cdot \tfrac{-i\nabla}{|i\nabla|}\psi}{|\nabla|\psi}\Big|\apprle \nlp{2}{\,|\nabla|^{3/2}\ph}\nlp{2}{\,|\nabla|^{1/2}\chi}=O(Yc^{-3/2}+Y^{3/2}\sqrt{\alpha}c^{-1}).
\]
We recall that $|g_0(p)-m(\alpha)|\le \min(\nlp{\infty}{g'_0} |p|,2\nlp{\infty}{g_0})$: it is $O(\min(1,\alpha |p|))$. Then we have:
\[
\begin{array}{rl}
\psh{g_0^2\psi}{|\nabla|\psi}&=m(\alpha)^2\psh{|\nabla|\psi}{\psi}+2m(\alpha)\psh{(g_0-m(\alpha))\psi}{|\nabla|\psi}+\psh{(g_0-m(\alpha))^2\psi}{|\nabla|\psi}\\
                                     &=m(\alpha)^2+O(\alpha Y^2),
 \end{array}
\]
Thus we have:
\[
\psh{|\D|^2\psi}{|\nabla|\psi}=m(\alpha)^2\psh{|\nabla|\psi}{\psi}+\psh{g_1^2|\nabla|\psi}{\psi}+O(\alpha Y^2).
\]

\noindent Let us now treat the term $\psh{\D(\mu-\alpha b_\g)\psi}{|\nabla|\psi}$ and first the term $\mu\psh{\D \psi}{|\nabla|\psi}$.

\[
\begin{array}{rl}
\psh{g_0\beta \psi}{|\nabla|\psi}&=\psh{g_0\psi}{|\nabla|\psi}-2\psh{g_0\chi}{|\nabla|\chi}=\psh{g_0\psi}{|\nabla|\psi}+O(\alpha Yc^{-2}+c^{-3})\\
                                   &=m(\alpha)\psh{|\nabla|\psi}{\psi}+O(\alpha Y^2+\alpha Y c^{-2}+c^{-3}),\\
\psh{\D \psi}{|\nabla|\psi}&=\psh{g_0\beta \psi}{|\nabla|\psi}+2\mathfrak{R}(\psh{S_\sigma\ph}{|\nabla|\chi})\\
                                &=m(\alpha)\psh{|\nabla|\psi}{\psi}+O(\alpha Y^2+\alpha Yc^{-2}+c^{-3}+Yc^{-3/2}+Y^{3/2}\sqrt{\alpha}c^{-1}),\\
\mu\psh{\D \psi}{|\nabla|\psi}&=m(\alpha)^2\psh{|\nabla|\psi}{\psi}+O(\alpha Y^2+Y^{3/2}\sqrt{\alpha}c^{-1}+Y c^{-3/2}+c^{-3}).
\end{array}
\]

We write:
\[
|d|^{1/2}R_\g\psi=\big[|d|^{1/2},R_\g\big]|d|^{-1} |d|\psi+R_\g|d|^{1/2}\psi,
\]
and thanks to Lemma \ref{rrq} we have: 
\[
\big\lVert\,\big[|d|^{1/2},R_\g\big]|d|^{-1}\big\rVert_{\mathfrak{S}_2}^2\apprle \nqu{\g}^2\apprle \nhi{\g}^2\apprle c^{-2}.
\]

By adapting the proof of Lemma \ref{chiant} we can prove the follwing estimates:
\[
\Big\lVert[|\nabla|,v_\g]\tfrac{1}{|\nabla||\D|^{1/2}}\Big\rVert_{\mathcal{B}},\Big\lVert [|\nabla|^{1/2},v_\g]\tfrac{1}{|\nabla|^{1/2}|\D|^{1/2}}\Big\rVert_{\mathcal{B}}\apprle  \ncc{\rho_\g}\sqrt{\log(\La)}.
\]
We use Lemmas \ref{bs6} and \ref{rrq} to get estimates of $\nlp{2}{b_\g \psi}.$ First we deal with the terms with $S_{\boldsymbol{\alpha}}$:
\[
\begin{array}{rl}
|\psh{R_\g\psi }{S_{\boldsymbol{\alpha}}|\nabla|\psi}|&\le  \big|\big\langle\big[|\nabla|^{1/2},R_\g\big] |d|^{-1}|d|\psi,\, S_{\boldsymbol{\alpha}}|\nabla|^{1/2}\psi\big\rangle\big|+ |\psh{R_\g |\nabla|^{1/2}\psi}{S_{\boldsymbol{\alpha}}|\nabla|^{1/2}\psi}|\\
       &\apprle Y \nqq{\g}(1+\nlp{2}{\nabla \psi})\apprle Y(L\alpha).
 \end{array}
 \]
The operator $S_{\boldsymbol{\alpha}}$ exchanges upper and lower spinors, so we get:
 \[
 \begin{array}{rl}
|\psh{S_{\sigma} v_\g\ph}{|\nabla|\chi}|&= \big|\psh{|\nabla| v_\g\ph}{S_{\sigma}\chi}\big|\le \big\lVert\,|\nabla| v_\g\ph\big\rVert_{L^2}\,\nlp{2}{S\chi}\\
                          &\le \alpha c^{-1}\Big\{ \big\lVert\big[\,|\nabla|, v_\g\big]\ph\big\rVert_{L^2}+\nlp{2}{v_\rho |\nabla|\ph}\Big\}\\
                          &\apprle \alpha c^{-1}\Big\{ \sqrt{\llo} \ncc{\rho_\g} \times \big\lVert\,|\nabla| |d|^{1/2} \ph\big\rVert_{L^2}+\ncc{\rho_\g}Y\Big\}\\
                          &\le Lc^{-5/2}(\nlp{2}{\nabla \ph}+\nlp{2}{\,|\nabla|^{3/2} \ph})\apprle Lc^{-7/2}+Lc^{-5/2}Y.
   \end{array}
   \]
   Similarly the following holds:
   \[
   \begin{array}{rl}
|\psh{S_{\sigma} v_\g\chi}{|\nabla|\ph}|&=\big|\psh{|\nabla|^{1/2} v_\g \chi}{|\nabla|^{1/2} S_{\sigma}\ph}\big| \\
        &\le \Big\{\big\lVert\big[|\nabla|^{1/2},v_\g\big] \chi\big\rVert_{L^2}+\nlp{2}{v_\g |\nabla|^{1/2} \chi}\Big\}\big\lVert\,|\nabla|^{3/2}\ph\big\rVert_{L^2}\\
        &\apprle \big(\sqrt{\llo}\ncc{\rho_\g}\nlp{2}{|\nabla|^{1/2} |d|^{1/2} \chi}+\ncc{\rho_\g}\big\lVert\,|\nabla|^{3/2}\chi\big\rVert_{L^2}\big)Y\\
        &\apprle (L\sqrt{\llo}c^{-1/2}(c^{-3/2}+c^{-1}\sqrt{\alpha Y} +\alpha c^{-2} Y))Y.
 \end{array}
 \]
 
\noindent We treat now the terms with $g_0(-i\nabla)$:
 \[
 \begin{array}{rllll}
 |\psh{v_\g\ph}{|\nabla|g_0\ph}|&\apprle& \ncc{\rho_\g}\big\lVert\,|\nabla|^{1/2}\ph\big\rVert_{L^2}\big\lVert\,|\nabla|g_0\ph\big\rVert_{L^2}&\apprle& Lc^{-2}\\
 |\psh{v_\g\chi}{|\nabla|g_0\chi}|&\apprle& \ncc{\rho_\g}\big\lVert\,|\nabla|^{1/2}\chi\big\rVert_{L^2}  \big\lVert\,|\nabla|g_0\chi\big\rVert_{L^2}&\apprle& Lc^{-2},\\
 |\psh{R_\g\psi}{|\nabla|g_0\psi}|&\apprle& \nqq{\g}\big\lVert\,|\nabla|^{1/2}\psi\big\rVert_{L^2} \big\lVert\,|\nabla|g_0\psi\big\rVert_{L^2} &\apprle &c^{-5/2}.
\end{array}
\]
It is clear that $\alpha(Lc^{-2}+c^{-5/2})=O(c^{-3})$. At last:

\[
\psh{\D(\mu-\alpha b_\g)\psi}{|\nabla|\psi}=m(\alpha)^2\psh{|\nabla|\psi}{\psi}+O(\alpha Y^2+Y^{3/2}\sqrt{\alpha}c^{-1}+Y c^{-3/2}+c^{-3}),
\]
and:
\[
Y^2(1-K\alpha)\le K_0c^{-3}+K_1(L\alpha^2)Y+K_3 \sqrt{\alpha}c^{-1}Y^{3/2}.
\]
As $\sqrt{\alpha} c^{-1}=O(c^{-3/4})$ (because $\alpha (\llo)^{1/4}=o(1)$ in the regime \eqref{regi}), we deduce $\psh{|\nabla|^{3}\psi}{\psi}=O(c^{-3})$, equivalently

\[\nso{3/2}{\unp}=O(1).\]

We now improve estimate \eqref{gros3} as written before:
\[
\begin{array}{rl}
g_0(p/c)-g_0(0)&=\dint_0^1 g'_0(tp/c) \frac{|p|}{c} dt=\dint_0^1(1-t) g''_0(tp/c) \frac{|p|^2}{c^2} dt\\
|g_0(p/c)-g_0(0)|^2&=\Big|\dint_0^1 g'_0(tp/c) dt\dint_0^1(1-u) g''_0(up/c) du\Big|\frac{|p|^3}{c^3},
\end{array}
\]
and therefore
\begin{equation}\nlp{2}{(m(\alpha)-\un{m})c\unp}\le K\sqrt{\frac{\lVert g'_0\rVert_{\infty}\,\lVert g''_0\rVert_\infty}{c}}=K\alpha\sqrt{L\alpha}=o(c^{-1}).
\end{equation}
So
\begin{equation}
\nso{1}{\un{\chi}}=O(c^{-1})\text{\ and\ }\nlp{2}{\,|\nabla|\chi}= O(c^{-2}).
\end{equation}

\subsubsection{Estimation of $E_{\text{BDF}}^0(1)$.}

Thanks to Eq.~\eqref{low}
\[
\chi=\frac{S_\sigma} {g_0+\mu}\ph+ \frac{\alpha}{g_0+\mu}\big((R_\g\psi)_2-v_\g\chi \big)=\frac{S_{\sigma}}{g_0+\mu}\ph+\delta\chi,
\]
where the remainder $\delta\chi$ is such that $\nlp{2}{\delta\chi}$ is lesser than 

\noindent$K\alpha (\nqq{\g}\nlp{2}{|\nabla|^{1/2}\psi}+\ncc{\rho}\nlp{2}{|\nabla|^{1/2}\chi})=O(\alpha c^{-3/2})=o(c^{-1})$. Thanks to Proposition \ref{g:estimates}, as $\nlp{2}{g_1\psi}=O(c^{-1})$,  we have the following asymptotic expansion:
\[
\begin{array}{rl}
E_{\text{BDF}}^0(1)+\tfrac{\alpha \WW(0)}{2c}D(\un{n},\un{n})&=\psh{g_0\ph}{\ph}-\psh{ \frac{g_0}{g_0+\mu}S_{\sigma}\ph}{\frac{1}{g_0+\mu}S_{\sigma}\ph}+2\mathfrak{R}\psh{\frac{1}{g_0+\mu}S_{\sigma}\ph}{S_{\sigma}\ph}+o(c^{-2})\\
         &= m(\alpha)(1-2\psh{\frac{g_1^2}{(g_0+\mu)^2}\ph}{\ph})+2\psh{\frac{g_1^2}{g_0+\mu}\ph}{\ph}+o(c^{-2})\\
         &=m(\alpha)-\psh{\frac{g_1^2}{2m(\alpha)}\ph}{\ph}+\psh{\frac{g_1^2}{m(\alpha)} \ph}{\ph}+o(c^{-2})\\
         &=m(\alpha)+\frac{1}{2m(\alpha)}\psh{g_1^2\ph}{\ph}+o(c^{-2})\\
         &=m(\alpha)+\frac{1}{2m(\alpha)}\psh{g_1^2\psi}{\psi}+o(c^{-2}).
\end{array}
\]
To deal with $g_0$ we use both results $\psh{|\nabla|^3\ph}{\ph}=O(c^{-3})$ and $|g'_0|=O(\alpha)$ and treat the $((g_0+\mu)^{-1})$'s one after the other. For the last line we use the fact that $\psh{|\nabla|^2\chi}{\chi}=O(c^{-3})$. Writing in terms of $\unp$:
\begin{equation}
C_0^2(E_{\text{BDF}}^0(1)-m(\alpha))=\frac{1}{(g'_1(0))^2(2\pi)^{3}}\dint c^2g_1\left(\frac{p}{c}\right)^2|\wh{\unp}(p)|^2dp-\diint \frac{|\unp(x)|^2|\unp(y)|^2}{|x-y|}dxdy+o(1).
\end{equation}

\noindent We recall (\emph{cf} Proposition \ref{vardg}, Appendix \ref{operateurD}) the $(g'_1)_{\alpha,\La}$'s are \emph{uniformly} continuous in a neighbourhood  of $0$; splitting in Fourier space at level $|p|=\sqrt{c}$ we get
\[
\begin{array}{l}
\underset{|p|\le \sqrt{c}}{\dint} c^2g_1(p/c)^2|\wh{\un{\psi}}(p)|^2dp=\dint\limits_{|p|\le\sqrt{c}} g'_1(0)^2|p|^2|\wh{\un{\psi}}(p)|^2dp\\
          \ \ \  \ \ \ \ \ \ \ \ \ +\dint\limits_{|p|\le\sqrt{c}} \left(\dint_{t=0}^1(g'_1(tp/c)-g'_1(0))dt \right)^2|p|^2|\wh{\un{\psi}}(p)|^2dp\\
          \ \ \ \ \ \  \ \ \ \ \ \ \ \ \ +2g'_1(0)\dint_{|p|\le \sqrt{c}} \left(\dint_{t=0}^1 (g'_1(tp/c)-g'_1(0))dt \right)|p|^2|\wh{\un{\psi}}(p)|^2dp\\
             \ \ \ \ \ \  =\dint\limits_{|p|\le\sqrt{c}} g'_1(0)^2|p|^2|\wh{\un{\psi}}(p)|^2dp+O\big( \lVert\,|\nabla|\un{\psi}\rVert ^2 \underset{|q|\le c^{-\tfrac{1}{2}}}{\sup}\big\{|g'_1(q)-g'_1(0)|\big\}\big)\\
             \ \ \ \ \ \   =\dint\limits_{|p|\le\sqrt{c}} g'_1(0)^2|p|^2|\wh{\un{\psi}}(p)|^2dp+\underset{c\to +\infty}{o}(1).
\end{array}
\]
Moreover:
\[
\begin{array}{rl}
\dint_{|p|\ge \sqrt{c}} c^2g_1(p/c)^2|\wh{\un{\psi}}(p)|^2dp &\apprle \dint_{|p|\ge \sqrt{c}} \frac{|p|^3}{|p|} |\wh{\un{\psi}}(p)|^2dp\\
                       &\apprle \frac{1}{\sqrt{c}}\psh{|\nabla|^3\un{\psi}}{\un{\psi}}\apprle c^{-1/2}\underset{c\to +\infty}{\to} 0.
\end{array}
\]
Thus
\[
\begin{array}{rl}
\dfrac{1}{(g'_1(0)^2)}\psh{c^2 g_1^2(\cdot/c)\un{\psi}}{\un{\psi}}-D(\un{n},\un{n})&=\psh{|\nabla|^2\un{\psi}}{\un{\psi}}-D(\un{n},\un{n})+o(1),
\end{array}
\]
By unicity of the asymptotic expansion and by definition of $E_{\text{CP}}$ we thus have
\begin{equation}
E_{\text{BDF}}^0(1)=m(\alpha)+C_0^{-2}E_{\text{CP}}+o((\alpha \WW(0))^2).
\end{equation}

As a consequence, the Choquard-Pekar energy wave function $\un{\psi}$ (more specifically $\un{\ph}$) tends to the minimizer. It is known \cite{L} there is but one minimizer in $H^1(\RR, \mathbf{C})$ up to translation. The fact that we work with spinors is harmless. By using convexity inequality for gradients \cite{LL} (Theorem 7.8 p.177) and Riesz's rearrangements inequality (sharp version in \cite{L}), we have that there is but one minimizer of the Choquard-Pekar energy in $H^1(\RR,\mathbf{C}^4)$ up to translation and \textit{overall} rotation in $\mathbf{C}^4$. Keeping track of the mass of $\un{\psi}$ with the help of some translation we get that necessarily it tends to a Choquard-Pekar minimizer.

\noindent\textit{Acknowledgment.} The author wishes to thank  \'E. S\'er\'e and M. Lewin for useful discussions and helpful comments, in particular the latter for suggesting the link with the Choquard-Pekar energy. He is also indebted to the referees and to \'E. Goujard for useful comments on the paper. This work was partially supported by the Grant ANR-10-BLAN 0101 of the French Ministry of Research.

\begin{appendices}
\section{The operator $\D$}\label{operateurD}
\begin{remark}
In this part the scalar product in $\mathbf{R}^d$ is written $\psh{\cdot}{\cdot}$ for $d=3,4$.
\end{remark}
\subsection{The functions $g_0$ and $g_1$}
As established in \cite{mf}, $\D$ is a solution to the following equation in the Fourier space
\begin{equation}\label{DD}
\wh{\D}=\wh{D^0}+\frac{\alpha}{4\pi^2}\wh{\frac{\D}{|\D|}}*\frac{1}{|\cdot|^2}\ \ \ \text{in}\ \mathcal{B}(B(0,\Lambda),\text{End}(\CC))
\end{equation}
and by a bootstrap argument $\wh{\D}\in\cap_{m\ge 1}H^m\big(B(0,\La)\big)$.
With the notation of \ref{gstar} (Subsection \ref{main}) it shows that $g_0,\mathbf{g_1}$ are smooth while $g_1(p)=\mathbf{g_1}(p)\cdot \om_p$ is \emph{a priori} in $\mathcal{C}^\infty(B(0,\Lambda)\backslash\{0\})$ and we have
\begin{subequations}
\begin{equation}\label{formg0}
g_0(|p|)=1+\frac{\alpha}{4\pi^2}\dint_{|r|<\Lambda}dr\frac{1}{|p-r|^2}\frac{g_0(|r|)}{\sqrt{g_1(|r|)^2+g_0(|r|)^2}},
\end{equation}
\begin{equation}\label{formg1}
g_1(|p|)=|p|+\frac{\alpha}{4\pi^2}\dint_{|r|<\Lambda}dr\frac{\langle\oo{p},\oo{r}\rangle}{|p-r|^2}\frac{g_1(|r|)}{\sqrt{g_1(|r|)^2+g_0(|r|)^2}}.
\end{equation}
\end{subequations}
\begin{remark}
We recall here that $C_1>0$ is a constant such that $g_1(r)\le C_1 r$ and $|g_0|_{\infty}\le C_1$.
\end{remark}

\begin{proposition}\label{g:estimates}
We have $g_1\in\mathcal{C}^1([0,\Lambda],\mathbf{R})$ and $g'_0(0)=0$. \newline Writing $\lVert\dd^2 g_1\rVert_\star =\underset{0<|p|\le\Lambda}{\sup}\big| \,|p|\dd^2 g_1(p)\big|$ the following holds:
\[
\left\{\begin{array}{l} \lVert g_0'\rVert_{\infty}=O(\alpha)\\ \lVert g_1'\rVert_{\infty}=O(1)\end{array}\right. \text{\ and\ }\left\{\begin{array}{l}  \lVert g_0''\rVert_{\infty}=O(\alpha)\\ \lVert\dd^2 g_1\rVert_\star=O(1) \end{array}.\right.
\]
Moreover there exists $K>0$ such that
\[
\forall\,q\in B(0,\La)\backslash B(0,1),\ \left\{\begin{array}{rl}|g_0(0)-1|&\le K\alpha\bigg\{\log\,\dfrac{\La}{|q|}+1\bigg\}\\ |g'_1(q)-1|&\le K\alpha\bigg\{\log\,\dfrac{\La}{|q|}+1\bigg\},\end{array}\right.
\]
and we have
\[
\begin{array}{ll}
g_0(0)=1+\dfrac{L}{\pi}+O(L^2+\alpha),&g'_1(0)=1+\dfrac{2L}{3\pi}+O(\alpha).
\end{array}
\]
\end{proposition}
In fact it suffices to differentiate \eqref{DD} to get $g'_0(p)$ and $g'_1(p)$, we take the norm to obtain the first part; then we differentiate once more to get the second part. The third part is a consequence of those parts.

\noindent Proposition \ref{g:estimates} enables us to prove the following result.
\begin{lemma}\label{m(p,q)}
Let $p,q\in B(0,\Lambda)$ and $k=p-q$. There holds \[\frac{\ed{p}\ed{q}-\langle\mathbf{g}(p),\mathbf{g}(q)\rangle}{\ed{p}\ed{q}(\ed{p}+\ed{q})}\le \min\Big(2,\dfrac{2K|k|^2}{\ed{p}^2},\dfrac{2K|k|^2}{\ed{q}^2}\Big).\]
where we can choose $K\le 2$ for $\alpha\llo$ sufficiently small.
\end{lemma}
\begin{dem}
In fact we can write for $a,b,t=b-a\in\RR$: $|a||b|-\langle a,b \rangle=\tfrac{a^2t^2-\langle t,a\rangle^2}{|a||b|+\langle a,b\rangle}$. If $\langle a,b\rangle>-\tfrac{|a||b|}{2}$ then $A=\tfrac{|a||b|-\langle a,b\rangle}{|a||b|}\le \tfrac{2a^2t^2}{a^2b^2}$, by symmetry we also have $A\le\tfrac{2b^2t^2}{a^2b^2}$. 

\noindent Else $-|a||b|\le \langle a,b\rangle\le -\tfrac{|a||b|}{2}$, then $\big\{|a||b|(|a||b|+\langle a,b\rangle)\big\}^{-1}\ge 2 (a^2b^2)^{-1}$, so:
\[
\begin{array}{l}
2\frac{t^2}{b^2}\ge 2\frac{a^2+b^2+|a||b|}{b^2}\ge 2\\
2\frac{t^2}{a^2}\ge 2\frac{a^2+b^2+|a||b|}{a^2}\ge 2.
\end{array}
\]
\end{dem}

\begin{proposition}\label{vardg} The function
\[
\dd \mathbf{g_1}(p)=\text{id}+\frac{\alpha}{4\pi^2}\int\limits_{|r|<\La} \frac{dr}{|p-r|\ed{r}}\Big(\dd \mathbf{g_1}(r)-\mathbf{g_1}(r)\frac{g_0(r)\dd g_0(r)+g_1(r)\dd g_1(r)}{\ed{r}^2}\Big)
\]
is in $\mathcal{C}^0(B(0,\Lambda),L(\RR,\CC))$ and 
\[|\dd \mathbf{g_1}(p)-\dd \mathbf{g_1}(q)|\le KL|p-q|.\]
In particular the same holds for $g_1'(t)=\langle \dd\mathbf{g_1}(t \om),\om \rangle$.
\end{proposition}

\noindent\textbf{Proof of Proposition \ref{g:estimates}}

1. We can define $\dd g_1(p)$ for $p\neq 0$. First we have
\[
\dd g_0(p) h= \frac{\alpha}{4\pi^2} \dint \frac{dq}{|p-q|^2}\left( \frac{\dd g_0(q)h}{\ed{q}}-\frac{g_0(q)\dd g_0(q)h+ g_1(q)\dd g_1(q)h}{\ed{q}^2}\frac{g_0(q)}{\ed{q}}\right).
\]

We remark that for $p\neq 0$ we have:
\[
\left\{
\begin{array}{l}
\dd g_\star(p)h= g'_\star(|p|)\psh{\om_p}{h},\ \star\in\{0,1\},\\
\psh{\dd \mathbf{g_1}(p)\cdot \om_p}{\om_p}=g'_1(|p|).
\end{array}
\right.
\]
Then
\[
\dd \mathbf{g_1}(p)\cdot h= h+\frac{\alpha}{4\pi^2}\dint \frac{dq}{|p-q|^2}\left( \frac{\dd \mathbf{g_1}(q)\cdot h}{\ed{q}}-\frac{g_0(q)\dd g_0(q)h+ g_1(q)\dd g_1(q)h}{\ed{q}^2}\frac{\mathbf{g_1}(q)}{\ed{q}}\right).
\]
So for any $\om\in \mathbf{S}^2$ we have
\begin{equation}\label{g1}
\begin{array}{rl}
g'_1(x)&=1+\dfrac{\alpha}{4\pi^2}\dint_{|q|\le\La} \frac{dq}{|x\om-q|^2}\bigg\{\Big(\frac{g_1(q)}{|q|}(1-\langle \om,\om_q\rangle^2)\Big)\frac{1}{\ed{q}}\\ 
&+\Big(g'_1(q) \langle\om_q,\om \rangle^2\big(1-\frac{g_1^2(q)}{\ed{q}^2}\big)\Big)\dfrac{1}{\ed{q}}-\dfrac{g_1(q)}{\ed{q}}\dfrac{\langle \om,\om_q\rangle^2}{\ed{q}}\dfrac{g_0(q)g'_0(q)}{\ed{q}}\bigg\}.
\end{array}
\end{equation}

The regularity of $g_1$ (as a function of $\mathbf{R}^+$) will come from the continuous extension to $x=0$ of the formula above.

We have
\begin{subequations}
\begin{equation}
|g'_0(|p|)|\le \frac{\alpha}{4\pi^2}\dint \frac{dq}{|p-q|^2}\left(\frac{|g'_0|_{\infty}}{\ed{q}}+|g_0|_{\infty}\frac{|g'_0|_{\infty}+|g'_1|_{\infty}}{\ed{q}^2}\right)
\end{equation}
\begin{equation}
|g'_1(|p|)|\le 1+\frac{\alpha}{4\pi^2}\dint \frac{dq}{|p-q|^2}\left(\frac{|g'_1|_{\infty}}{\ed{q}}+ \frac{|g'_0|_{\infty}+|g'_1|_{\infty}}{\ed{q}}\right).
\end{equation}
\end{subequations}
Thus
\[
\left\{
\begin{array}{l}
|g'_0|_{\infty}\le K_1 \alpha \llo |g'_0|_{\infty}+ K_2\alpha |g'_1|_{\infty}\\
|g'_1|_{\infty}\le 1+K_3\alpha \llo  (|g'_0|_{\infty}+|g'_1|_{\infty}).
\end{array}
\right.
\]
So $|g'_0|_{\infty}\apprle \alpha$ and $|g'_1|_{\infty}\le 1+K\alpha \llo$.

\noindent Since $g_0\in\mathcal{C}^\infty(B(0,\Lambda),\mathbf{R})$ \emph{and} radial, necessarily  
\[\dd g_0(0)=0\text{\ and\ }g_0'(0)=\dd g_0(0)\,\om=0,\,\forall \om\in\mathbf{S}^2.\]

2. We treat now the second derivative $\dd^2 \D$. We write $h_\star=\tfrac{g_\star}{\ed{\cdot}}$

\noindent and $\mathcal{J}=\ed{\cdot}^{-1}$. The coefficient of $\beta$ in $\dd^2\D(p)\,h^2$ is
\[
\dd^2 g_0(p)\, h^2=\frac{\alpha}{4\pi^2}\dint_q \frac{dq}{|p-q|^2} \dd^2 h_0(q)\,h^2,
\]
where
\[
\begin{array}{rl}
\dd^2 h_0(q)\, h^2&=\dfrac{\dd^2 g_0(p)\cdot h^2}{\ed{q}}-\dfrac{2}{\ed{q}^3}\dd g_0(q)h\,[g_0(q)\dd g_0(q)h+g_1(q)\dd g_1(q)h]\\
 &\ -\dfrac{g_0(q)}{\ed{q}^3}\left[(\dd g_0(q)h)^2+g_0(q)\dd ^2g_0(q)h^2 +(\dd g_1(q)h)^2+g_1(q)\dd ^2g_1(q)h^2 \right]\\
 &\ \ +3\dfrac{g_0(q)}{\ed{q}^5}[g_0(q)\dd g_0(q)h+g_1(q)\dd g_1(q)h]^2.
\end{array}
\]

Furthermore, we have
\[
\dd^2 \mathbf{g}_1(p)h^2=\dfrac{\alpha}{4\pi^2} \dint \frac{dq}{|p-q|^2} \bigg(\frac{\dd^2 \mathbf{g}_1(q)h^2}{\ed{q}}+2\dd \mathbf{g_1}(q)h \,\dd \mathcal{J}(q)h+\mathbf{g_1}(q)\dd ^2 \mathcal{J}(q)h^2\bigg).
\]
Since we have $ \langle |p| \dd^2 \mathbf{g_1}(p) h^2,\om_p\rangle= |p|\dd^2 g_1^p\cdot h^2+\frac{g_1(p)}{|p|}\left(\langle \om_p,h \rangle^2-|h|^2 \right)$, by taking the scalar product with $\oo{p}$ we get
\[
\begin{array}{rl}
|p| |\dd^2 g_1(p)| &\le C_1+\dfrac{\alpha}{4\pi^2}\dint \frac{|p|dq}{|p-q|^2|q|\ee{q}}\lVert \dd^2 g_1\rVert_\star +\frac{\alpha}{4\pi^2}\dint \frac{|p|dq}{|p-q|^2\ee{q}^2}\lVert\dd^2 g_1\rVert_\star\\
                &+\dfrac{\alpha}{4\pi^2}\dint_q\frac{|p|dq}{|p-q|^2}\bigg\{\frac{1}{\ee{q}^2}(|\dd g_0|^2+|\dd g_1|^2)+\frac{g_0(q)}{\ee{q}^2}|\dd^2 g_0|\\
                &+\dfrac{3}{\ee{q}^2}(|\dd g_0|+|\dd g_1|)^2+2(|\dd g_1|+C_1)\dfrac{|\dd g_0|+|\dd g_1|}{\ee{q}^2}+\dfrac{1}{\ee{q}}\dfrac{2|\dd g_1|+4C_1}{|q|}\bigg\}.
\end{array}
\]

We also have:
\[
\begin{array}{rl}
|\dd^2 g_0(p)|&\le \dfrac{\alpha}{4\pi^2}\bigg\{\dint\frac{C_1dq}{\ee{q}^2|p-q|^2}\lVert\dd^2 g_1\rVert_\star\\
                             &\ \ \ \ \ \ \ \ \ \dint_q\frac{dq}{|p-q|^2}\Big(\frac{|\dd^2 g_0|}{\ee{q}}+2\frac{|\dd g_0|(|\dd g_0|+|\dd g_1|)}{\ee{q}^2}+\frac{g_0(q)}{\ee{q}}\frac{|\dd g_0|^2+|\dd g_1|^2}{\ee{q}^2}\\
                             &\ \ \ \ \ \ \ \ \ +\dfrac{g_0(q)^2}{\ee{q}^2}\dfrac{|\dd^2 g_0|}{\ee{q}}+3\dfrac{g_0(q)}{\ee{q}}\dfrac{(|\dd g_0|+|\dd g_1|)^2}{\ee{q}^2}\Big)\bigg\}.
\end{array}
\]

As $\tfrac{|p|}{|p-q|^2|q|}\le2\max(\tfrac{1}{|p-q||q|},\tfrac{1}{|p-q|^2})$, we have
\[
\dint_{|q|\le\La} \frac{dq |p|}{|p-q|^2 |q|\ee{q}}\le 2\left(\dint_{|q|\le\La}\frac{dq}{|p-q||q|\ee{q}}+\dint_{|q|\le\La}\frac{dq}{|p-q|\ee{q}}\right).
\]
We recall then that the convolution of radial nonnegative functions is radial nonnegative. So the following holds:
\[
\left\{\begin{array}{ll} \lVert g''_0\rVert_{\infty}\le K\alpha\\ \lVert\dd^2 g_1\rVert_\star\le C_1+K\alpha\llo \end{array}\right.
\]

3. By Ineq~\eqref{formg0} and for $p\in\RR$, $1\le |p|<\La$ we get that:
\[
\begin{array}{rl}
\dfrac{4\pi^2|g_0(p)-1|}{\alpha}&=\underset{|q|<\La}{\dint}\dfrac{dq}{|p-q|^2}\dfrac{g_0(q)}{\sqrt{g_0(x)^2+g_1(q)^2}}\le\underset{|q|<\La}{\dint}\dfrac{dq}{|p-q|^2}\dfrac{1}{\sqrt{1+\tfrac{|q|^2}{|g_0|_{\infty}^2}}}\\
              &\le \underset{|q|<2\La}{\dint}\dfrac{dq}{|q|^2}\dfrac{1}{\sqrt{1+\tfrac{|p+q|^2}{|g_0|_{\infty}^2}}}\le  \underset{|q|<2\La}{\dint}\dfrac{dq}{|q|^2}\dfrac{|g_0|_{\infty}}{|p+q|}\\
              &=2\pi|g_0|_{\infty}\dint_0^{\La}\dfrac{dr}{r|p|}\log\,\bigg|\frac{r+|p|}{r-|p|}\bigg|=2\pi|g_0|_{\infty}\dint_0^\La\frac{r+|p|-\big|r-|p|\big|}{r|p|}\\
              &\apprle 1+\log\,\dfrac{\La}{|p|}.
\end{array}
\]
To deal with $g'_1$ we use Eq.~\eqref{g1}. The integral of the integrand in the second line is $O(1)$: as we multiply by $\alpha$ its contribution is $O(\alpha)$. For $1\le |p|<\La$ there holds:
\[
\underset{|q|<\La}{\dint} \frac{dq}{|p-q|^2|\ed{q}}\frac{g_1(q)}{|q|}\le \underset{|q|<2\La}{\dint} \frac{dq}{|q|^2|p+q|}\apprle 1+\log\,\dfrac{\La}{|p|}.
\]
For $g_0(0)$ we have:
\[
\begin{array}{rl}
\dfrac{\pi|g_0(0)-1|}{\alpha}&=\dint_0^\La \frac{g_0(r)dr}{\sqrt{g_0(r)^2+g_1(r)^2}}=\dint_1^\La dr\bigg\{\frac{1+O(\alpha\log\,\tfrac{\La}{r})}{\sqrt{1+r^2}}\bigg\}+O(1)\\
     &=\log(\La)+O(1+\alpha\log(\La)^2).
\end{array}
\]
Let us prove the estimation of $g'_1(0)$. There holds for any $0<x<\La$ and $\omega \in\mathbf{S}^2$:
\[
\begin{array}{rl}
\underset{|q|<\La}{\dint}\dfrac{\langle \oo,\oo{q}\rangle^2dq}{|x\omega-q|^2\ed{q}}\dfrac{g_1(q)}{|q|}&=2\pi\dint_0^{\La}dr\frac{x^2+r^2}{2x^2}\bigg[\frac{x^2+r^2}{2rx}\log\,\Big|\frac{x+r}{x-r}\Big|-1\bigg]\frac{g_1(r)}{r\ed{r}},\\
     &=2\pi \dint_0^{\La/x}dr\frac{1+r^2}{2r}\bigg[\frac{1+r^2}{2r}\log\,\Big|\frac{1+r}{1-r}\Big|-1\bigg]\frac{g_1(xr)}{\ed{xr}}.
\end{array}
\]
We split at two levels: $e^{-1}$ and $e$. The integral over $(e^{-1},e)$ is $O(1)$ for $\log$ is integrable on $(0,e)$. For $x\in(0,e^{-1})$ there holds the following expansion:
\[
\frac{1+r^2}{r}(\log(1+r)-\log(1-r))-1=\frac{4}{3}r^2+\underset{r\to 0}{O}(r^3),
\]
thus the integration over $(0,e^{-1})$ is $O(1)$. For $x\in(e,\La/x)$ there holds:
\[
\frac{1+r^2}{r}(\log(1+r^{-1})-\log(1-r^{-1}))-1=\frac{4}{3r^2}+\underset{r\to+\infty}{O}(r^{-3}).
\]
If we multiply by $\dfrac{1+r^2}{2r}$ we get $\dfrac{2}{3r}+\underset{r\to+\infty}{O}(r^{-2})$. Thus the integration over $(e,\La)$ gives:
\[
\frac{4\pi}{3}\dint_e^{\La/x}\frac{g_1(rx)dr}{\ed{rx}r}+O(1)=\frac{4\pi}{3}\dint_{ex}^{\La}\frac{g_1(r)dr}{\ed{r}r}+O(1).
\]
At last we get:
\[
g'_1(0)-1=\frac{\alpha}{\pi}\dint_0^{\La}\frac{g_1(r)dr}{r\ed{r}}\Big[1-\frac{1}{3}\Big]+O(\alpha)=\frac{2\alpha\llo}{3\pi}+O(\alpha).
\]
\hfill{\footnotesize$\Box$}

\noindent\textbf{Proof of Proposition \ref{vardg}}
In fact it suffices to use another formulae for $\dd^2 \mathbf{g}_1$ and $\dd^2 g_0$ consisting in replacing $g_1(q)\dd g_1(q)$ by
\[
\psh{\mathbf{g}_1(q)}{\dd \mathbf{g}_1(q)}.
\]
By the same method as for $\dd g_0,\dd g_1$, we get that
\begin{equation}
\big\lVert  \dd^2 \mathbf{g}_1   \big\rVert_{\infty}\apprle L.
\end{equation}
%to split $B(0,\La)$ in two domains: 

%\noindent We write $F_p=\RR\cap\{r:|p-r|\le |q-r|\},\,F_q=\RR\cap\{r:|q-r|\le |p-r|\}.$

%In $F_p\cap B(0,\La)$ we take spherical coordinates with respect to $p$, in $F_q\cap B(0,\La)$ with respect to $q$. 
%There holds 
%\[
%\big||p-r|^{-1}-|q-r|^{-1}\big|\le\left\{ \begin{array}{rl}
%\frac{|p-q|}{|p-r|^2} & \text{\ for\ }r\in F_p,\\
%\frac{|p-q|}{|q-r|^2} & \text{\ for\ }r\in F_q.
%\end{array}\right.
%\]
%For $p,q\in B(0,\La)$ we have
%\begin{equation}
%\int_{|l|<\La}\big||p-l|^{-1}-|q-l|^{-1}\big|\frac{dl}{\ed{l}}\le 8\pi|p-q|\dint_{r=-\La}^{\La} \frac{dr}{\sqrt{1+r^2}}\apprle \llo|p-q|.
%\end{equation}
\hfill{\footnotesize$\Box$}

\subsection{The function $B_\La$}
We recall that 
\begin{equation}\label{integrand}
B_\La(k)=\frac{1}{\pi^2|k|^2}\dint\limits_{|p=l-\frac{k}{2}|,|q=l+\frac{k}{2}|<\La}\frac{\ed{p}\ed{q}-\langle\mathbf{g}(p),\mathbf{g}(q)\rangle}{\ed{p}\ed{q}(\ed{p}+\ed{q})}dl\ge 0.
\end{equation}
This formula holds only for $k\neq 0$: our first purpose is to extend it continuously to $0$. Thanks to Lemma \ref{m(p,q)} we can say that $B_\La(k)\le K\llo.$

\begin{notation}
Throughout this part, $p=\ell+\tfrac{k}{2}, q=\ell-\tfrac{k}{2}$.
\end{notation}

\begin{proposition}\label{bla}
Let $\om$ be any in $\mathbf{S}^2$. For $\ell\in B(0,\La)$ we write:
\[
\mathbf{g}_\ell^\om:=\begin{pmatrix}g'_0(|\ell|)\oo{\ell}\cdot\om\\\dd \mathbf{g_1}(\ell)\cdot \om\end{pmatrix}\text{\ and\ }\widetilde{E}^\om_\ell:=|\mathbf{g}_\ell^\om|.
\]
Then we have
\begin{equation}\label{intform}
B_\La(k)\underset{k\to 0}{\to} \frac{1}{\pi^2}\dint_{|\ell|\le\La}\frac{|\mathbf{g}_\ell^\om\wedge\mathbf{g}_\ell|^2}{4\ed{\ell}^5}d\ell=:B_\La(0),
\end{equation}

Moreover
\[
B_\La(0)=\frac{2}{3\pi}\llo+O(L\llo+1).
\]
\end{proposition}

\begin{dem}
Let us write $I=\pi^2|k|^2B_\La(k),$ its integrand $f(\ell)$ and $x=|k|$. Let us consider $0<\eps<\tfrac{2}{3}$ and $s=\tfrac{1}{3}+\eps$. We assume $x<1$ and split the domain in three:
\[
\begin{array}{l}
B=\{\ell:\,|\ell|\le x^s\}, A=\{\ell:\,x^s<|\ell|< \La-\frac{x}{2}\},\\
C=\{\ell\,:\,|\ell-\frac{k}{2}|,|\ell+\frac{k}{2}|<\La\}\backslash\{\ell\,:\,|\ell|<\La-\frac{x}{2}\}\subset \{\ell\,:\,\La-\frac{x}{2}<|\ell|<\La\}=C'.
\end{array}
\]
Using Lemma \ref{m(p,q)} we get the following behaviour \emph{independent} of $\alpha,\La$ in the regime \eqref{regi}:
\begin{equation}
|I_B|\le Kx^{2+3s}=Kx^{3+3\eps}=\underset{x\to 0}{o}(x^3),\ \ |I_C|\le Kx^2\log\left(\frac{\La}{\La-\tfrac{x}{2}}\right)\underset{x\to 0}{\sim}\frac{Kx^3}{\La}.
\end{equation}
There remains to deal with $I_A$: we rewrite $f(\ell)$ as follows:
\begin{equation}
f(\ell)=\frac{|\mathbf{g}(p)\wedge \mathbf{g}(q)|^2}{\ed{p}\ed{q}(\ed{p}+\ed{q})(\ed{p}\ed{q}+\mathbf{g}(p)\cdot \mathbf{g}(q))}
\end{equation}
where $|\mathbf{g}(p)\wedge \mathbf{g}(q)|^2=\sum_{i}|\Delta_{0i}|^2+\sum_{i,j}|\Delta_{ij}|^2$,
\begin{subequations}\label{mineur}
\begin{equation}
\Delta_{0i}=\bigg|\begin{matrix} g_0(p) & g_0(q)\\ (\mathbf{g_1}(p))_i & (\mathbf{g_1}(q))_i\end{matrix}\bigg|= \bigg|\begin{matrix} \delta g_0 & g_0(q) \\ (\delta \mathbf{g_1})_i & (\mathbf{g_1}(q))_i\end{matrix}\bigg|
\end{equation}
\begin{equation}
\Delta_{ij}=\bigg|\begin{matrix} (\mathbf{g_1}(p))_i & (\mathbf{g_1}(q))_i\\ (\mathbf{g_1}(p))_j & (\mathbf{g_1}(q))_j\end{matrix}\bigg|=\bigg|\begin{matrix} (\delta \mathbf{g_1})_i & (\mathbf{g_1}(q))_i \\ (\delta\mathbf{g_1})_j & (\mathbf{g_1}(q))_j\end{matrix}\bigg|
\end{equation}
\end{subequations}
$\delta g_\star=g_\star(p)-g_\star(q)$.

If we take $k$ along a \emph{fixed} half-line: $k=x\om$ we have
\[
\begin{array}{rl}
\frac{1}{x}\delta g_0(k,\ell) &=\dint_{t=0}^1 \dd g_0(\ell+(t-1/2)k)\cdot \om dt\underset{x\to 0}{\to} g'_0(|\ell|) \oo{\ell}\cdot \om\\
\frac{1}{x}\delta \mathbf{g_1}(k,\ell) &=\dint_{t=0}^1 \dd \mathbf{g_1}(\ell+(t-1/2)k)\cdot \om dt\underset{x\to 0}{\to} \dd \mathbf{g_1}(\ell)\cdot \om.
\end{array}
\]

In fact, as $A,g_0,g_1$ are radial symmetrics so is $I_A(k)$ and for $\om\in\mathbf{S}^2$ \emph{fixed} and $p'=\ell+\tfrac{x\om}{2}$, $q'=\ell-\tfrac{x\om}{2}$ there holds
\[
I_A(k=x\om_k)=\frac{1}{\pi^2x^2}\dint\limits_{x^s<|\ell|<\La-\frac{x}{2}}\frac{\ed{p'}\ed{q'}-\langle\mathbf{g}(p'),\mathbf{g}(q')\rangle}{\ed{p'}\ed{q'}(\ed{p'}+\ed{q'})}dl,
\]
$f_0(\ell)=\tfrac{f(\ell)}{x^2}\chi_{\ell\in A}$ is also symmetric. By Proposition \ref{g:estimates} we have

\noindent $|f_0(\ell)|\le K\tfrac{1}{(1+|\ell|^2)^{3/2}}\chi_{|\ell|\le\La-x/2}$. By dominated convergence we get the integral formula \eqref{intform}.

\noindent As there holds by symmetry
\begin{equation}
\dint_{\mathbf{n}\in \mathbf{S}^2} \langle \mathbf{n},\om\rangle^2 d\mathbf{n}=\frac{4}{3}\pi,\ \ \ 
\dint_{\mathbf{n}\in \mathbf{S}^2} |\dd \mathbf{g_1}(|\ell|\mathbf{n})\cdot \om|^2d\mathbf{n}=\frac{4}{3}\pi \left((g'_1)^2(\ell) +2\frac{g_1(\ell)^2}{|\ell|^2}\right)
\end{equation}
we have
\[
\begin{array}{l}
B_\La(0)=\frac{1}{3\pi}\bigg( \dint_{u=0}^\La u^2\frac{((g'_0)^2(u)+(g_1')^2(u)+2\tfrac{g_1(u)^2}{|u|^2})(g_0^2(u)+g_1^2(u))}{(g_0(u)^2+g_1(u)^2)^{5/2}}du\\
\ \ \ \ \ \ \ \ \ \ \ \ \ \ \ \ \ \ \ \ \ \ \ \ \ \ \ \ \ -\dint_{u=0}^\La u^2\frac{(g_0g'_0(u)+g_1g'_1(u))^2}{(g_0(u)^2+g_1(u)^2)^{5/2}}du\bigg),
\end{array}
\]

and

\begin{equation*}
B_\La(0)=\frac{1}{3\pi}\left( \dint_{u=0}^\La u^2\frac{ (g'_0)^2(u)+(g_1')^2(u)+2\tfrac{g_1(|u|)^2}{|u|^2}}{(g_0(u)^2+g_1(u)^2)^{3/2}}du-\dint_{u=0}^\La u^2\frac{(g_0g'_0(u)+g_1g'_1(u))^2}{(g_0(u)^2+g_1(u)^2)^{5/2}}du\right).
\end{equation*}
Thanks to Proposition \ref{g:estimates}, we get the estimate of $B_\La(0)$.
\end{dem}

\noindent Let us look at the variations $|k|^{-1}|B_\La(k)-B_\La(0)|$. 
\begin{proposition}\label{z(x)} There exists $0<r_\eps\in\mathbf{R}^+$, \emph{independent} of $\alpha,\La$ in the regime \eqref{regi} such that 

\noindent for $|k|<r_\eps$:
\[|k|^{-1}|B_\La(k)-B_\La(0)|\le K(\La^{-1}+L^2|k|+|k|^{3\eps}+|k|^{2/3-\eps}).\]
Choosing $\eps:=6^{-1}$ there holds:
\[
|k|^{-1}|B_\La(k)-B_\La(0)|\le K(\La^{-1}+|k|^{1/2}).
\]
\end{proposition}
\begin{dem}

For $k\in B(0,1)$ we write $|k|=x$. We reconsider the domains $A,B$ and $C$ of the proof of Proposition \ref{bla} and write $f_1$ the integrand in \eqref{integrand}.

We have $|\int_B f_1|\le Kx^{3s}=\underset{x\to 0}{O}(x^{1+3\eps})$ and $|\int_C f_1|\le K\log(\tfrac{\La}{\La-x/2})=\underset{x\to 0}{O}(\tfrac{x}{\La})$. There remains the integration over $A$. For $|\ell|\ge x^s$ we have $\tfrac{x}{|\ell|}=O(x^{2/3-\eps})$ so we can expand the integrand of $I_A(x)$ at order $1$. Indeed:
\[
\ed{p}^{-1}=\ed{\ell}^{-1}\Big\{1+\frac{\ed{p}-\ed{\ell}}{\ed{\ell}}\Big\}^{-1}=\ed{\ell}^{-1}\Big\{1+\frac{\ed{\ell}-\ed{p}}{\ed{\ell}}+\underset{x\to 0}{O}\big(\frac{x^2}{\ed{\ell}^2}\big)\Big\},
\]
\noindent where the $\underset{x\to 0}{O}(\cdot)$ is independent of $\ell$ (because $\ed{\ell}\ge 1$) . The same holds for $\ed{q}^{-1}$ and $(\ed{p}+\ed{q})^{-1}$. Writing $h(\ell,k)=\ed{p}\ed{q}-\mathbf{g}(p)\cdot\mathbf{g}(q)$ we have:
\[
\begin{array}{rl}
I_A(x)=&\dfrac{1}{x^2}\dint_A \frac{h(\ell,k)}{2\ed{\ell}^3}d\ell+\frac{1}{x^2}\dint_A\frac{h(\ell,k)}{2\ed{\ell}^3}\Big(\frac{2\ed{\ell}-\ed{p}-\ed{q}}{\ed{\ell}}\\&+\dfrac{2\ed{\ell}-\ed{p}-\ed{q}}{2\ed{\ell}}+O\big(\dfrac{x^2}{\ed{\ell}^2}\big)\Big).
\end{array}
\]
By Taylor formula (at order $2$):

\[|2\ed{\ell}-(\ed{p}+\ed{q})|\le \int_t\int_u dtdu K x^{1+2/3-\eps}=K x^{1+2/3-\eps}.\]

\noindent By Proposition \ref{vardg} and by Taylor formula at order $1$ we have:
\[
\bigg|\frac{\mathbf{g}(p)-\mathbf{g}(q)}{x}-\mathbf{g}^\om_l \bigg|\apprle Lx.
\]
Thus $|k|^{-1}|B_\La(k)-B_\La(0)|=\underset{k\to 0}{O}(\La^{-1}+L+|k|^{3\eps})$.
\end{dem}

\section{The fixed point method: estimations}\label{estptf}

\subsection{Estimation about the $R_\cdot$ operator}
Let us generalize Lemma 8.\cite{ptf} that states the inequality: $\nqr{R_Q}\apprle \nq{Q}$. Further generalisations are detailed in \cite{these}.
\begin{lemma}\label{change8}
Let $f$ be some function $f:B(0,\La)\to\mathbf{R}_+$ and $Q\in\mathcal{Q}_f$. Then we have:
\begin{equation}
\diint f(p-q)\frac{|\wh{R}_Q(p,q)|^2}{|p+q|}dpdq\apprle \diint f(p-q)|p+q||\wh{Q}(p,q)|^2dpdq.
\end{equation}
\end{lemma}
\begin{dem}
The kernel $\wh{R}(p,q):=\wh{R}_Q(p,q)$ is equal to:
\begin{equation*}
\wh{R}(p,q)=\frac{1}{2\pi^2}\dint \frac{\wh{Q}(p-\ell,q-\ell)}{|\ell|^2}d\ell.
\end{equation*}
We remark the Fourier multiplier:
\[
A(x,y)\mapsto \mathscr{F}^{-1}\left\{f(p-q) \wh{A}(p,q)\right\}
\]
commutes with $R_\cdot: A\mapsto R_A$. So it suffices to show that:
\[
\diint\frac{|\wh{R}(p,q)|^2}{|p+q|}dpdq\apprle \diint |p+q||\wh{Q}(p,q)|^2dpdq.
\]
To this end we follow the proof in \cite{ptf}, for any $\theta\in (0,2)$:
\[
\begin{array}{rl}
 \diint\frac{|\wh{R}(p,q)|^2}{|p+q|}dpdq&=8\diint \frac{dudv}{|2u|}|\wh{R}(u+v,u-v)|^2\\
      & \le \frac{8}{(2\pi^2)^2}\diiiint \frac{|\wh{Q}(\ell+v,\ell-v)||\wh{Q}(\ell'+v,\ell'-v)|}{|2u| |\ell-u|^2|\ell'-u|^2}dudvd\ell d\ell'\\
      & \le \frac{8}{(2\pi^2)^2}\diiiint \frac{1}{|2u|}\frac{|\wh{Q}(\ell+v,\ell-v)|^2}{|\ell-u|^2|\ell'-u|^2}\frac{|2\ell|^{1+\theta}}{|2\ell'|^{1+\theta}}dudvd\ell d\ell'\\
      &\le \frac{8}{(2\pi^2)^2}\diint |2\ell||\wh{Q}(\ell+v,\ell-v)|^2w_\theta(\ell)dvd\ell,
     \end{array}
\]
where the weight $w_\theta(\ell)$ is:
\[
 w_\theta(\ell):=|2\ell|^\theta\diint \frac{dud\ell'}{|2u||2\ell'|^{1+\theta}|\ell-u|^2|\ell'-u|^2}.
\]
Then we have:
\[
\begin{array}{rl}
 w_\theta(\ell)&\le \dint_u \frac{du}{|2u|^{1+\theta}|u-\ell|^2}\Big(|2u|^\theta\dint_{\ell'}\frac{d\ell'}{|2\ell'|^{1+\theta}|\ell'-u|^2}\Big)\\
 &\le \Big(\dfrac{1}{2}\dint \frac{dx}{|x|^{1+\theta}|x-\mathbf{e}|^2}\Big)^2,
\end{array}
\]
where $\mathbf{e}\in\mathbf{R}^3$ is any vector satisfying $|\mathbf{e}|=1$.
\end{dem}

\subsection{Estimates for the fixed point method}\label{needed}

Let $N_0\ge 0$ be in $\mathfrak{S}_1(\hl)$ and let $\g_0$ be in $\mathfrak{S}_1^{\PP}(\hl)$. 

\noindent We write $n_0:=\rho_{N_0}$ and $x(N_0):=\ns{2}{\nabla N_0}$. We assume that
\begin{equation}
 \ttr(N_0)\apprle 1
\end{equation}
to simplify. In our problem $N_0=\ket{\psi}\bra{\psi}$ with $\nlp{2}{\psi}=1$.

%(
In this part $f$ is some function $f:\RR\mapsto [1,+\infty)$ satisfying condition \eqref{conditions} and we consider the Fourier multiplier $m_f$:
\[
 Q(x,y)\in L^2(\hl\times\hl)\mapsto \mathscr{F}^{-1}(f(p-q)\wh{Q}(p,q)).
\]
%]

For $Q_0\in\mathcal{Q}_f,\rho_0\in\mathfrak{C}_f$ we write:
\[
\nxf{(Q_0,\rho_0)}:=K_{(0)}(f)(\nqf{Q_0}+\ncf{\rho_0}),
\]
where $K_{(0)}(f)>0$ to be precised later.

\noindent By Kato's inequality and Sobolev  inequality~\eqref{sobin} $\ncc{n_0}\apprle x^{1/2}$ and $\nlp{2}{n_0}\apprle x^{3/2}$. For the last inequality it suffices to write $N_0:=\sum a_i \ket{f_i}\bra{f_i}$, $a_i\ge 0$ and $\nlp{2}{f_i}=1$. Then:
\[
 \nlp{2}{n_0}\le \sum_i a_i \nlp{2}{\nabla f_i}^{3/2}\apprle (\sum_i a_i \nlp{2}{\nabla f_i}^2)^{3/4}.
\]
The same method enables us to prove that $\ns{2}{R_{N_0}}\apprle x$.

\begin{lemma}\label{lin}
Let $N_0$ and $\g_0$ be as above. Then we have:
\[
\begin{array}{rl}
\nqf{Q_{0,1}[\rho_{\g_0}]}&\apprle \sqrt{\llo}\ncf{\rho_{\g_0}},\\
\nqf{Q_{1,0}[\g_0]}&\apprle \nqf{ \g_0},\\
\ncf{\rho_{1,0}[\rho_{\g_0}]}&\apprle \sqrt{\llo}\nqf{ \g_0}.
\end{array}
\]
Moreover:
\[
 \begin{array}{rl}
  \nhi{Q_{1,0}[N_0]}&\apprle x,\\
  \ncc{\rho_{1,0}[N_0]}&\apprle x.
 \end{array}
\]

\end{lemma}

\begin{lemma}\label{nonlin}
Let $(Q_0,\rho_0)$ be in $\mathcal{X}_f$. There exist constants $\text{K}_{(1)},\text{K}_{(2)}>0$ such that, writing
\[
 G_f(Q,\rho):=\text{K}_{(1)}C(f)(\nqf{Q}+\ncf{\rho})
\]
we have:
\begin{equation}
 \forall \ell\ge 2:\ \nxf{(Q_\ell,\rho_\ell)[Q_0,\rho_0]}\le \frac{\text{K}_{(2)}}{\sqrt{\ell}}G_f(Q_0,\rho_0)^{\ell}.
\end{equation}

\end{lemma}

Assuming these lemmas hold, we follow \cite{ptf} to find a ball $B(0,R_f)$ invariant under the function $F=F_Q\times F_\rho$ of the fixed point method (\eqref{cauchy1} and \eqref{cauchy2}) and on which $F$ is a contraction. Indeed for some $K_{(4)}>0$, we have:
\[
\left\{
\begin{array}{l}
\nqf{F_Q[Q_0,\rho_0]}\le \nqf{N}+K_{(4)}\sqrt{L\alpha}(\nqf{Q_0}+\ncf{\rho_0})+K_{(2)}\ssum_{\ell=2}^{+\infty}\ell^{1/2}(\alpha G_f(Q_0,\rho_0))^\ell,\\
\ncf{F_\rho[Q_0,\rho_0]}\le \ncf{n}+K_{(4)}\sqrt{L\alpha}(\nqf{Q_0}+\ncf{\rho_0})+K_{(2)}\ssum_{\ell=2}^{+\infty}\ell^{1/2}(\alpha G_f(Q_0,\rho_0))^\ell,
\end{array}
\right.
\]
these upper bounds are finite provided $\alpha G_f(Q_0,\rho_0)<1$ where $G_f$ is defined in Lemma \ref{nonlin}. Moreover:
\[
\lVert \text{d} F[Q_0,\rho_0] \rVert_{\text{L}(\mathcal{X}_f)}\le 2\big\{K_{(4)}\sqrt{L\alpha}+\alpha K_{(3)}(f)\ssum_{\ell=2}^{+\infty}\ell^{3/2}(\alpha G_f(Q_0,\rho_0))^{\ell-1}\big\}
\]
where $K_{(3)}(f)=K_{(1)}K_{(2)}C(f)A(f)$. The supremum of the above upper bound on $B_{\mathcal{X}_f}(0,R)$ is written $\nu=\nu(f,R)$.

We take $K_{(0)}(f):=K_{(1)}C(f)$, $R_f=\eps_f\sqrt{\llo}$ for some $\eps_f>0$ and assume $(\nqf{N}+\ncf{n})\le \eps_n\sqrt{\llo}$ (with $0<\eps_n<\eps_f$). 

For any $(Q_0,\rho_0)\in B_{\mathcal{X}_f}(0,R_f)$ the following holds:
\[
\begin{array}{rl}
\nxf{F(Q_0,\rho_0)}&\le \nu(f,R_f) \nxf{(Q_0,\rho_0)}+\nxf{F(0,0)}\\
                                   &\le \nu(f,R_f) \eps_f\sqrt{\llo}+K_{(0)}(f)\eps_n \sqrt{\llo}.
 \end{array}
\]
We have:
\[
\nu(f,\eps_f \sqrt{\llo})\le 2K_{(4)}\sqrt{L\alpha}+2\alpha K_{(3)}(f)\ssum_{\ell=2}^{+\infty} \ell^{1/2}\bigg(\dfrac{\alpha \eps_f\sqrt{\llo}}{C(f)}\bigg)^{\ell-1}
\]
To apply the Banach fixed point Theorem it suffices to have:
\[
\nu(f,\eps_f \sqrt{\llo})<1\ \text{and}\ \dfrac{\nu(f,\eps_f\sqrt{\llo})+K_{(1)}C(f)\eps_n}{\eps_f}< 1.
\]

For $f_j(p-q)=\ed{p-q}^j$ with $j\in\{0,1,2\}$ and provided $\alpha\sqrt{\llo}\eps_f$ is small enough we have:
\[
\nu(f_1,\eps_{f} \sqrt{\llo})\apprle \sqrt{L\alpha}(1+\alpha \sqrt{\llo}\eps_f)=O(\sqrt{L\alpha}).
\]

In the case $\alpha \llo\ll 1$, it suffices to take $\tfrac{\eps_n}{\eps_f}$ small enough to apply the fixed point Theorem.

\noindent\textbf{Proof of Lemma \ref{lin}}
Let $M(\cdot,\cdot)$ be the function 
\[
(p,q)\in B(0,\La)^2\mapsto M(p,q):=\frac{1}{\ed{p}+\ed{q}}\Big(\frac{\wh{\D}(p)}{\ed{p}}\frac{\wh{\D}(q)}{\ed{q}}-1\Big). 
\]
We write $S(p):=\tfrac{\wh{\D}(p)}{\ed{p}}$ for short. A direct computation in Fourier space (and Cauchy's formula) gives like in \cite{ptf}:
\begin{equation}\label{ftQ}
\left\{
 \begin{array}{rl}
 \wh{Q}_{0,1}(\rho;p,q)&=\frac{1}{2^{5/2}\pi^{3/2}}\wh{\ph}_\rho(p-q)M(p,q),\\
 \wh{Q}_{1,0}(\g;p,q)&=-\frac{1}{2}\big(S(p)\wh{R}_\g(p,q) S(q)-\wh{R}_\g(p,q)\big).
 \end{array}
 \right.
\end{equation}
We will use Lemma \ref{m(p,q)}: it gives an estimation of $M(p,q):$
\[
 |M(p,q)|\apprle \frac{|p-q|}{(\ed{p}+\ed{q})^2}.
\]
The estimation of $\nqf{Q_{0,1}(\rho_{\g_0})}$ is then easy. In \eqref{ftQ}, it suffices to use Lemma \ref{change8} to get estimation of $\nqf{Q_{1,0}(\g_0)}$.

Then, as $\ns{2}{R_N}\apprle x$, the estimation of $\nqkin{Q_{1,0}(N_0)}$ follows from a simple computation of $\diint |\wh{Q}_{1,0}(N_0)|^2$.

Then the norm $\ncf{\rho_{1,0}[\g_0]}$ is dealt with in the same way as in \cite{ptf}:
\[
\rho_{1,0}[\g_0;k]=-\frac{1}{2^{5/2}\pi^{3/2}}\underset{|u\pm\tfrac{k}{2}|<\La}{\dint}\ttr_{\mathbf{C}^4}\Big\{\wh{R}_{\g_0}\big(u+\frac{k}{2},u-\frac{k}{2}\big) M\big(u-\frac{k}{2},u+\frac{k}{2}\big)\Big\}.
\]
By Cauchy-Schwartz inequality we have:
\[
\begin{array}{l}
|\rho_{1,0}[\g_0;k]|^2\le \dfrac{1}{2^5\pi^3}\dint \ed{2u}^{-1}|\wh{R}_{\g_0}(u+\tfrac{k}{2},u-\tfrac{k}{2})|^2du\ \times\\
\ \ \ \ \ \ \ \ \ \ \ \ \ \ \ \ \ \ \dint\ed{2u}|M(u-\tfrac{k}{2},u+\tfrac{k}{2})|^2du.
\end{array}
\]
By Lemma \ref{change8} $\ncf{\rho_{1,0}[\g_0]}\apprle \sqrt{\llo}\nqf{\g_0}$.

\hfill{\footnotesize$\Box$}

\noindent\textbf{Proof of Lemma \ref{nonlin}}
We only sketch the proof of Lemma \ref{nonlin} in this paper: we refer the reader to \cite{ptf,these} for full details. 

The main idea is to use the K.-S.-S.  inequality~\eqref{K.-S.-S.} together with the Hölder inequality for $\mathfrak{S}_p(\hl)$. 
For instance, let us take the Hilbert-Schmidt norm of $Q_{0,3}[\rho_0]$: writing $h_{\rho_0}:=\mathscr{F}^{-1}(|\wh{\ph}_{\rho_0}|)$ we have
\[
\big\lVert\dfrac{1}{(|\D|^2+\eta^2)^{1/4}} h_{\rho_0}\dfrac{1}{(|\D|^2+\eta^2)^{1/4}}\big\rVert_{\mathfrak{S}_6}\apprle \frac{1}{\ed{\eta}^{1/2}}\nlp{6}{h_{\rho_0}}.
\]
By Sobolev inequality we get $\nlp{6}{h_{\rho_0}}\apprle \ncc{\rho_0}$ and thus there holds:
\[
\begin{array}{rl}
 \ns{2}{Q_{0,3}[\rho_0]}&\le \dfrac{1}{2\pi}\dint_{-\infty}^{+\infty}\big\lVert\dfrac{1}{(|\D|^2+\eta^2)^{1/4}} h_{\rho_0}\dfrac{1}{(|\D|^2+\eta^2)^{1/4}}\big\rVert_{\mathfrak{S}_6}^3\frac{d\eta}{\ed{\eta}}\\
                      &\apprle \ncc{\rho_0}^3.
 \end{array}
\]

Let us first estimate $\nqf{Q_\ell},\ \ell\ge 2$.

The term $Q_{2}$ is dealt with the same way as in \cite{ptf}: we refer to this paper for details.

The difference between the example above $\ns{2}{Q_{0,3}[\rho_0]}$ and $\nqf{Q_\ell[Q_0,\rho_0]}$ is that we have to multiply $\wh{Q}_\ell(p,q)$ by the weight
\[
 \sqrt{f(p-q)\ed{p+q}}
\]
before taking the Hilbert-Schmidt norm. Besides this fact the main idea is the same:
\begin{itemize}
 \item We consider $ \sqrt{f(p-q)\ed{p+q}}\wh{Q}_\ell$,
 \item We take its Hilbert-Schmidt norm and get an upper bound of it using K.-S.-S. and Hölder inequalities.
\end{itemize}

To deal with $\sqrt{f(p-q)}$ we use condition \eqref{conditions}:
\begin{equation}\label{trickf}
 \sqrt{f(p-q)}\le C(f)^{\ell-1}\big\{\sqrt{f(p-p_1)}+\sqrt{f(p_1-p_2)}+\cdots+\sqrt{f(p_{\ell-1}-q)}\big\}
\end{equation}
and to deal with $\sqrt{\ed{p+q}}$ we use the following trick:
\begin{equation}\label{trickkin}
 \dfrac{1}{\{(\ed{p}^2+\eta^2)(\ed{q}^2+\eta^2)\}^{1/4}}\apprle \dfrac{1}{\sqrt{\ed{p+q}\ed{\eta}}}.
\end{equation}

We consider the integral representation of each term of $\wh{Q}_{j,\ell-j}[Q_0,\rho_0]$; for convenience we write $R_0:=R[Q_0]$ and $\ph_0:=\ph[\rho_0]$.

For instance let us treat the term where the $j$ operators $R_0$ are on the left, we take the modulus and get the upper bound:
\begin{equation}\label{int_chiant}
\begin{array}{l}
\dint_{-\infty}^{+\infty}\frac{d\eta}{(2\pi)^{1+3(\ell-j)/2}}\underset{B(0,\La)^{\ell-1}}{\dint} dp_1 \cdots d p_{\ell-1}\frac{|\wh{R}_0(p,p_1)|}{\sqrt{\ed{p}^2+\eta^2}}\displaystyle\prod_{i=1}^j \frac{|\wh{R}_0(p_i,p_{i+1})|}{\sqrt{\ed{p_i}^2+\eta^2}}\times\\
\ \ \ \ \ \ \ \ \ \ \ \ \ \ \ \ \ \ \ \ \ \ \ \displaystyle\prod_{k=j+1}^{\ell-1}\frac{|\wh{\ph}_0(p_k-p_{k+1})|}{\sqrt{\ed{p_{k+1}}^2+\eta^2}}.
\end{array}
\end{equation}
We write $p_0:=p$ and $p_\ell:=q$.

We multiply \eqref{int_chiant} by $\sqrt{f(p-q)\ed{p+q}}$ and use tricks \eqref{trickf} and \eqref{trickkin}. We then use \eqref{trickkin} for the terms involving $p_i$ and $p_{i+1}$ ($0\le i\le j-1$) and get:
\begin{equation}
 \begin{array}{rl}
\big\lVert f(p'-q')^{1/2} |\wh{R}_0(p',q')|/\sqrt{\ed{p'+q'}}\big\rVert_{\mathfrak{S}_2}&\apprle \nqf{Q_0}.
 \end{array}
\end{equation}

Moreover we have by the K.-S.-S. inequality:
\begin{equation}\label{s6phi}
 \begin{array}{l}
  \big\lVert(\ed{p'}^2+\eta^2)^{-1/4} |\sqrt{f(p'-q')}\wh{\ph}_0(p'-q')|  (\ed{q'}^2+\eta^2)^{-1/4}\big\rVert_{\mathfrak{S}_6}\apprle \dfrac{\ncf{\rho_0}}{\ed{\eta}^{1/2}},\\
  \big\lVert(\ed{p'}^2+\eta^2)^{-1/4} |\sqrt{f(p'-q')}\wh{\ph}_0(p'-q')|  (\ed{q'}^2+\eta^2)^{-1/4}\big\rVert_{\mathfrak{S}_{\infty}}\\
  \ \ \ \ \ \ \ \ \ \ \ \ \ \ \ \ \ \ \ \ \ \ \le \big\lVert(\ed{p'}^2+\eta^2)^{-1/4} |\sqrt{f(p'-q')}\wh{\ph}_0(p'-q')|  (\ed{q'}^2+\eta^2)^{-1/4}\big\rVert_{\mathfrak{S}_6}.
 \end{array}
\end{equation}

By using those K.-S.-S. inequalities under the integral sign $\int_\eta$ in \eqref{int_chiant} (multiplied by the weight $\sqrt{f(p-q)\ed{p+q}}$), we get an upper bound of the form: 
\[
\dint_{-\infty}^{+\infty}\frac{d\eta}{\ed{\eta}^{(1+j+\ell-j)/2}}\ell K^\ell C(f)^\ell\nqf{Q_0}^j\ncf{\rho_0}^{\ell-j}.
\]
This upper bound is valid provided $(\ell+1)/2> 1$ and $\ell\ge 3$ \textit{ie} if $\ell\ge 3$. 

In fact the same method gives:
\[
\begin{array}{rl}
 \nqf{Q_{2,0}[Q_0]}&\apprle C(f)^2\nqf{Q_0}^2,\\
 \nqf{Q_{1,1}[Q_0,\rho_0]}&\apprle C(f)^2\nqf{Q_0}\ncf{\rho_0}.
\end{array}
\]

Let us now deal with the densities $\rho_\ell[Q_0,\rho_0]$. First remark: as recalled in \cite{ptf}, Furry's Theorem states that for all $\ell=2\ell_1\in 2\mathbf{N}^*$ \emph{even}, we have
\[
\rho_{0,2\ell_1}=0. 
\]
As in \cite{ptf}, we deal with the other terms \emph{by duality}: the dual $\mathfrak{C}_f'$ of $\mathfrak{C}_f$ is:
\[
 \mathfrak{C}_f'=\Big\{ \zeta\in\mathcal{S}'(\mathbf{R}^3)\,:\ \dint\frac{|\wh{\zeta(k)}|^2}{|k|f(k)}dk<+\infty\Big\}.
\]
For any $\zeta\in \mathfrak{C}_f'\cap L^2$ and $Q\in\mathfrak{S}_2(\hl)$ we have
\[
 Q\zeta=(Q|\D|^2) (\tfrac{1}{|\D|^2} \zeta)\in\mathfrak{S}_1(L^2(\mathbf{R}^3)).
\]
Above, it is understood that $\tfrac{1}{|\D|^2}$ is the Fourier multiplier
\[
 \dfrac{1}{|\D|^2}:\phi\in L^2(\mathbf{R}^3)\mapsto \mathscr{F}^{-1}\big\{\frac{\chi_{|p|<\La}}{\ed{p}^2}\wh{\phi}(p)\big\}\in L^2(\mathbf{R}^3).
\]
Then the following holds:
\[
|\psh{\rho_Q}{\zeta}|=|\ttr(Q\zeta)|=|\ttr(\wh{Q\zeta})|\le \dint |\wh{Q\zeta}(p,p)|dp.
\]
The idea is to get an upper bound depending only on the $\mathfrak{C}'_f$-norm of $\zeta$ and to conclude by density of $\mathfrak{C}_f'\cap L^2$ in $\mathfrak{C}_f'$.

The ingredients are the same but we treat $\rho_{1,1}$ and $\rho_{0,3}$ differently (as in \cite{ptf}). We use the same K.-S.-S. inequalities and \eqref{trickf}.

For instance, for $\ell\ge 5$:
\[
 \begin{array}{l}
 |\wh{Q_{0,5}\zeta}(p,p)|\le \dfrac{1}{(2\pi)^{1+3(\ell+1)/2}}\dint_{-\infty}^{+\infty}\underset{B(0,\La)^{\ell-1}}{\dint d\mathbf{p}} \frac{|\wh{\ph}_0(p-p_1)|}{(\ed{p}^2+\eta^2)^{1/4}}\displaystyle\prod_{k=1}^{\ell-1}\frac{|\wh{\ph}_0(p_k-p_{k+1})|}{\sqrt{\ed{p_k}^2+\eta^2}}\times\\
 \ \ \  \ \ \  \ \ \  \ \ \  \ \ \  \ \ \  \ \ \  \ \ \ \dfrac{|\wh{\zeta}(p_{\ell-1}-p)|}{(\ed{p}^2+\eta^2)^{1/4}}.
 \end{array}
\]
We write
\[
 |\wh{\zeta}(p_{\ell-1}-p)|=\frac{\sqrt{f(p_{\ell-1}-p)}}{\sqrt{f(p_{\ell-1}-p)}}|\wh{\zeta}(p_{\ell-1}-p)|=\sqrt{f(p_{\ell-1}-p)}\wh{\zeta'}(p_{\ell-1}-p)
\]
and use \eqref{trickf}:
\[
 \sqrt{f(p_{\ell-1}-p)}\le C(f)^{\ell-1}(\sqrt{f(p_1-p)}+\cdots+\sqrt{f(p_{\ell-1}-p_{\ell-2})}).
\]
Then it suffices to use $6$ times the first inequality of \eqref{s6phi} and $(\ell+1-6)$ times the second.

We refer the reader to \cite{ptf,these} for the other terms.
\hfill{\footnotesize$\Box$}

\subsection{Estimates of a fixed point}\label{estimfp}
Let $(N,n)\in\mathcal{X}_\star$ be given where $\star$ means $0,1$ or no subscript. Let us assume that the norms of $N$ and $n$ are $O(1)$ such that we can apply the fixed point Theorem (\emph{cf} Lemmas \ref{fpq} and \ref{fpqu}). From now on $\nu$ is Lipschitz constant in Lemma \ref{fpq} that is the one corresponding to $F$ applied on some ball $B_{\mathcal{X}}(0,R)$. We write: $x=\sqrt{\ttr(-\Delta |N|)}$.

We apply the Banach theorem with initial data $(0,0)\in\mathcal{X}_\star:$ iterations are written $(\g'_{(\ell)},\rho'_{(\ell)})$ and $\g_{(\ell)},\rho_{(\ell)}$ are defined as follows:
\begin{equation}
\g'_{(\ell)}=\g_{(\ell)}+N,\ \rho'_{(\ell)}=\rho_{(\ell)}+n
\end{equation} with $\g_{(\ell+1)}=\chi_{(-\infty,0)}(\D+\alpha(\ph_{\rho'_{(\ell)}}-R(\g_{(\ell)}')))-\PP$. The fixed point is written:
\[
(\g',\rho'_\g)=(\g,\rho_\g)+(N,n).
\]

\begin{lemma}
Let $N,n,\g,\rho_\g$ be as above. If $\lVert(N,n)\rVert_{\mathcal{X}_\star}=O(1)$ then so is $\lVert(\g,\rho_\g)\rVert_{\mathcal{X}_\star}$. 
\end{lemma}
\begin{dem}
In the regime \eqref{regi}, the Lipschitz constant $\nu_0$ in Lemmas \ref{fpq}, \ref{fpqu} is $o(1)$. So:
\[
 \begin{array}{rl}
  \lVert(\g',\rho_\g')-(0,0)\rVert_{\mathcal{X}_\star}&\le \ssum_{\ell=0}^{+\infty}\lVert(\g_{(\ell+1)}',\rho_{\g_{(\ell+1)}}')-(\g_{(\ell)}',\rho_{\g_{(\ell)}}')\rVert_{\mathcal{X}_\star}\\
                  &\le \ssum_{\ell=0}^{+\infty}\nu_0^{\ell}\lVert(\g_{(1)}',\rho_{\g_{(1)}}')-(\g_{(0)}',\rho_{\g_{(0)}}')\rVert_{\mathcal{X}_\star}\\
                  &\le \frac{\lVert F(0,0)\rVert_{\mathcal{X}_\star}}{1-\nu_0}\le \frac{\lVert(N,n)\rVert_{\mathcal{X}_\star}}{1-\nu_0}.
 \end{array}
\]
\end{dem}

We want to be more precise and prove Lemma \ref{florilege_estimations}. We first have:
\begin{lemma}\label{noir}
Let $N,n,\g,\rho_\g,x(N)=:x$ be as above. Let us write:
\[
w(N):=\sqrt{\diint |p-q|^2 |p+q| |\wh{N}(p,q)|^2dpdq}.
\]
Then the following estimates hold:
\[
\begin{array}{rl | rl}
\nhi{\g}&\apprle \sqrt{L\alpha}x^{1/2}+\alpha x+L\alpha,& \ncc{\rho_\g}&\apprle Lx^{-1/2}+\alpha x+L\alpha,\\
\nq{\g}&\apprle  \sqrt{L\alpha}x^{1/2}+\alpha,&\nc{\rho_\g}& \apprle Lx^{-1/2}+\alpha x+w(N)\sqrt{L\alpha}+L\alpha.
\end{array}
\]
\end{lemma}

\textbf{Proof:} The first point is devoted to Lemma \ref{noir} and the second to the end of Lemma \ref{florilege_estimations}.

\noindent 1.\ We write $\ov{n}:=\mathscr{F}^{-1}(\wh{n}(k)/(1+\alpha B_\La(k)))$. There holds: $F(0,0)=(N,\ov{n})$; in particular $\g_{(1)}=0$ and $\rho_{(1)}=\ov{n}-n=-\mathscr{F}^{-1}(\WW)*n$. 

Writing $\g=\sum_{\ell=1}^{+\infty} (\g_{(\ell+1)}-\g_{(\ell)})+\g_{(1)}$ we have:
\[
\begin{array}{rl}
\nhi{\g}&\le \sum_{\ell=2}^{+\infty} \nq{\g_{(\ell+1)}-\g_{(\ell)}}+\nhi{\g_{(2)}-\g_{(1)}}+\nhi{\g_{(1)}}\\
             &\le \sum_{\ell=2}^{+\infty}\nu^\ell\nxf{F(0,0)}+\nhi{F_Q(N,\ov{n})-N}.
\end{array}
\]
The first term on the right hand side is equal to $\tfrac{\nu^2}{1-\nu}\nxf{N,\ov{n}}=O(L\alpha)$. The second term is the $\nhi{\cdot}$ norm of:
\[
\ssum_{j=1}^{+\infty}\alpha^j Q_j[N,\ov{n}].
\]
By Lemmas \ref{lin} and \ref{nonlin} the following inequalities hold: $\alpha \nhi{Q_1[N,\ov{n}]}\apprle \sqrt{L\alpha} x^{-1/2}+\alpha x$ and 
\[
\sum_{j=2}^{+\infty}\alpha^j\nhi{Q_j[N,\ov{n}]}\le \sum_{j=2}^{+\infty}\alpha^j\nq{Q_j[N,\ov{n}]}\apprle \alpha^2 \nx{(N,\ov{n})}=O(\alpha^2)=O(L\alpha).
\]

Using the same method for $\nq{\cdot}$, we have $\alpha \nq{Q_1[N,\ov{n}]}\apprle \sqrt{L\alpha} x^{-1/2}+\alpha \nq{N}$ and:
\begin{equation}\label{nhinq}
\begin{array}{rl}
\nhi{\g}&\apprle \sqrt{L\alpha} x^{-1/2}+\alpha x+L\alpha\apprle \sqrt{L\alpha} x^{-1/2}+\alpha x+L\alpha,\\
\nq{\g}&\apprle \sqrt{L\alpha} x^{-1/2}+\alpha \nq{N}+L\alpha\apprle  \sqrt{L\alpha} x^{-1/2}+\alpha.
\end{array}
\end{equation}
There remains to check that $x=O(L\alpha)$ to get $\nhi{\g}\apprle L\alpha$ and $\nq{\g}\apprle \alpha$.

For the density we have:
\[
\begin{array}{rl}
\ncc{\rho_\g}&\le \ssum_{\ell=2}^{+\infty} \nc{\rho_{(\ell+1)}-\rho_{(\ell)}}+\ncc{\rho_{(2)}-\rho_{(1)}}+\ncc{\rho_{(1)}}\\
             &\le \ssum_{\ell=2}^{+\infty}\nu^\ell\nxf{F(0,0)}+\ncc{F_\rho(N,\ov{n})-\ov{n}}+\ncc{n-\ov{n}}.
\end{array}
\]
It is clear that $\ncc{n-\ov{n}}\le \nlp{\infty}{\WW}\ncc{n}\apprle Lx^{-1/2}$. The first term is $O(\alpha^2).$ The last term is the norm of:
\[
\mathscr{F}^{-1}\bigg\{\frac{1}{1+\alpha B_\La(k)}\Big(\alpha \wh{\rho}_{1,0}(N,k)+\ssum_{j=2}^{+\infty} \alpha^j \wh{\rho}_j(N,\ov{n};k)\Big)\bigg\}.
\]
We use Lemmas \ref{lin} and \ref{nonlin} to get:
\[
\begin{array}{rl}
\alpha \ncc{(\delta_0-\WW)*\rho_{1,0}(N)}&\apprle \alpha x,\\
\ssum_{j=2}^{+\infty}\alpha^j \nc{(\delta_0-\WW)*\rho_j(N,\ov{n})}&\apprle \alpha^2\nx{(N,\ov{n})}^2\apprle \alpha^2\apprle L\sqrt{L\alpha}.
\end{array}
\]
Here $\delta_0$ is the usual Dirac's generalized function.

If we consider the norm $\nc{\cdot}$, there holds:
\[
\alpha \nlp{2}{(\delta_0-\WW)*\rho_{1,0}(N)}\apprle \sqrt{L\alpha}\sqrt{\diint |p+q||p-q|^2 |\wh{N}(p,q)|^2dpdq}=:\sqrt{L\alpha}w(N)
\]
where we have used Lemma \ref{change8} with $f(p-q)=|p-q|^2$. Provided $x=O(L\alpha)$ and $w(N)=O(L)$ the following estimate hold:
\[
\nc{\rho_\g},\ncc{\rho_\g}\apprle L\sqrt{L\alpha}.
\]

For the test function defined by \eqref{testn} and \eqref{testg}, it is clear that $x=O(L\alpha)$ and $w(N)=O((L\alpha)^{3/2})$.

\noindent 2.\ The estimate of $\ns{2}{\g}$ follows from these estimates. First by computing in Fourier space it is clear that:
\[
\forall \rho_0\in\mathcal{C}:\ \ns{2}{Q_{0,1}[\rho_0]}\apprle \ncc{\rho_0}.
\]
Then:
\[
\begin{array}{rl}
\ns{2}{\g}&\le \ssum_{j=1}^{+\infty}\alpha^j \ns{2}{Q_j[\g',\rho'_\g]}\\
                &\le \alpha (\ns{2}{Q_{0,1}[\rho_\g']}+\ns{2}{Q_{1,0}[\g']})+\ssum_{j=2}^{+\infty}\alpha^j \nhi{Q_j[\g',\rho'_\g]}\\
                &\apprle \alpha \ncc{\rho'_\g}+\alpha(\ns{2}{R_N}+\nqq{\g})+O(\alpha^2\nx{(\g',\rho'_\g)}^2)\apprle \alpha\sqrt{L\alpha}.
\end{array}
\]

Moreover:
\[
\diint |\ed{p}-\ed{q}|^2|\wh{\g}(p,q)|^2dpdq\apprle \diint |p-q|^2|\wh{\g}(p,q)|^2dpdq\apprle \nhi{\g}^2\apprle (L\alpha)^2.
\]

To conclude this part, there remains to estimate $\nlp{2}{\g|\D| \psi_\la}$ and $\nlp{2}{\g \psi_\la}$. We have:
\[
\begin{array}{rl}
\nlp{2}{\g|\D| \psi_\la}&\le \ns{2}{\g}\nlp{2}{\,|\D| \psi_\la}\apprle \alpha\sqrt{L\alpha},\\
\nlp{2}{\g \psi_\la}&\le  \ns{2}{\g}\nlp{2}{\psi_\la}\apprle \alpha\sqrt{L\alpha}.
\end{array}
\]
We can get better upper bounds \cite{these} but we do not need them here.\hfill{\footnotesize$\Box$}

\end{appendices}

\end{document}